\documentclass[sigplan,twocolumn,nonacm]{acmart}
\usepackage[OT1]{fontenc}
\usepackage{textcomp}
\usepackage{booktabs} 
\usepackage{textcomp}
\usepackage{booktabs} % For formal tables
\usepackage{comment}
\usepackage{wrapfig}
\usepackage{xcolor}
\usepackage{graphicx}
\usepackage{subfig}
\usepackage{array}
\usepackage{multicol}
\usepackage{afterpage}
\usepackage{float}
\usepackage{listings}
\usepackage{multirow}
\usepackage{arydshln}
\usepackage{xspace}
\usepackage{enumitem}
\usepackage{fancyhdr}
\usepackage{amsmath}
\usepackage{url}
\usepackage[htt]{hyphenat}
\usepackage{subfloat}
\usepackage{amsfonts}

\usepackage{cleveref}
\crefformat{chapter}{\S#2#1#3}
\crefformat{section}{\S#2#1#3} 
\crefformat{subsection}{\S#2#1#3}
\crefformat{subsubsection}{\S#2#1#3}

\graphicspath{{figures/}}

\usepackage{pifont}
\usepackage{xcolor}

\newcommand{\phm}[1]{\vspace{.4em} \noindent\textbf{#1}\hspace{.5em}}

\usepackage{array}  

\usepackage{booktabs}

\usepackage{xcolor}
\usepackage[linesnumbered,ruled,vlined]{algorithm2e}

\begin{abstract}
Applications based on Large Language Models (LLMs) contains a series of tasks to address real-world problems with boosted capability, which have dynamic demand volumes on diverse backends. Existing serving systems treat the resource demands of LLM applications as a blackbox, compromising end-to-end efficiency due to improper queuing order and backend warm up latency. We find that the resource demands of LLM applications can be modeled in a general and accurate manner with \emph{Probabilistic Demand Graph} (PDGraph).
We then propose Hermes, which leverages PDGraph for efficient serving of LLM applications.
Confronting probabilistic demand description, Hermes applies the Gittins policy to determine the scheduling order that can minimize the average application completion time.
It also uses the PDGraph model to help prewarm cold backends at proper moments.
Experiments with diverse LLM applications confirm that Hermes can effectively improve the application serving efficiency, reducing the average completion time by over 70\% and the P95 completion time by over 80\%.  

\end{abstract}

\begin{document}

%% Title information
%\title{Hermes: Towards Coinference-Centric Serving for LLM Applications} %Correlation-aware LLM Inference Serving}         %% [Short Title] is optional;
\title{Efficient Serving of LLM Applications with Probabilistic Demand Modeling}         %% [Short Title] is optional;
% \title{Hermes: Efficient Serving of LLM Application with Correlation-aware Demand Modeling}
%\title{Hermes: Correlation-aware Request Scheduling for LLM Inference Workloads}         %% [Short Title] is optional;
                                        %% when present, will be used in
                                        %% header instead of Full Title.

	  \author{Yifei Liu}
		\affiliation{%
			\institution{Shanghai Jiao Tong University}
                \country{China}
		}

        \author{Zuo Gan}
		\affiliation{%
			\institution{Shanghai Jiao Tong University}
                \country{China}
		}

        \author{Zhenghao Gan}
		\affiliation{%
		\institution{Shanghai Jiao Tong University}
                \country{China}
		}

         \author{Weiye Wang}
		\affiliation{%
		\institution{Shanghai Jiao Tong University}
                \country{China}
		}

         \author{Chen Chen}
		\affiliation{%
		\institution{Shanghai Jiao Tong University}
                \country{China}
		}
        \authornote{Chen Chen is the corresponding author.}

         \author{Yizhou Shan}
		\affiliation{%
		\institution{Huawei Cloud}
                \country{China}
		}

         \author{Xusheng Chen}
		\affiliation{%
		\institution{Huawei Cloud}
                \country{China}
		}

         \author{Zhenhua Han}
		\affiliation{%
		\institution{Unaffiliated}
                \country{China}
		}
        
         \author{Yifei Zhu}
		\affiliation{%
		\institution{Shanghai Jiao Tong University}
                \country{China}
		}
        
         \author{Shixuan Sun}
		\affiliation{%
		\institution{Shanghai Jiao Tong University}
                \country{China}
		}
        
         \author{Minyi Guo}
		\affiliation{%
		\institution{Shanghai Jiao Tong University}
                \country{China}
		}

\maketitle

\pagestyle{plain}

%% Keywords
%% comma separated list
% \keywords{keyword1, keyword2, keyword3}  %% \keywords are mandatory in final camera-ready submission

%% \maketitle
%% Note: \maketitle command must come after title commands, author
%% commands, abstract environment, Computing Classification System
%% environment and commands, and keywords command.

%!TEX root = main.tex
\section{Introduction}
\label{sec:intro}

Large Language Models (LLMs)~\cite{brown2020language,floridi2020gpt,touvron2023llama} have demonstrated its strong capability in language understanding and generation. 
% been widely adopted in various domains~\cite{dosovitskiy2020image,stiennon2020learning,chen2021evaluating}.
%Yet, a single LLM request is often deficient for real-world tasks. 
Yet, even though LLMs are evolving rapidly~\cite{sonoda2024diagnostic,reid2024gemini,liu2024deepseek}, it is commonly recognized that a single LLM request is often deficient for many real-world problems~\cite{compound_ai}. 
For example, the input context windows of typical LLMs are often of limited sizes (e.g., less than 10M tokens)~\cite{chen2023extending,lin2024infinite}, thus processing a large document would require issuing multiple parallel inference requests. 
Meanwhile, the output of an LLM request may be unreliable (i.e., suffering \emph{hallucination}~\cite{ji2023survey}), and additional LLM requests would be required to ensure output quality (e.g., with self-reflection~\cite{ji2023towards}).
Moreover, the built-in knowledge and interaction modality of LLMs are also limited, and non-LLM tasks like docker execution~\cite{factool_code,docker_execution} or third-party tool calling~\cite{shen2024llm,shen2024hugginggpt} are often integrated to augment LLM capabilities~\cite{abhyankarinfercept}.
% For example, AI agents may repeatedly submit an LLM request for multiple times to mitigate hallucinations~\cite{ji2023survey}, and LLM requests may be augmented with external requests (e.g., knowledge retrieval~\cite{lewis2020retrieval}, code testing~\cite{factool_code}, tool calling~\cite{abhyankarinfercept,shen2024llm}) to possess the capability to handle diverse tasks.
We call such a set of correlated LLM and non-LLM tasks---which collaborate to address a realistic problem---as an \emph{LLM application}.
LLM applications would be a mainstream AI workload paradigm in the future. % becoming a mainstream workload paradigm to exploit LLM capability in reality.  

LLM applications are often hosted on the cloud~\cite{assistants_api,ibne2024ensuring,lin2024parrot}, and it becomes critical to serve them efficiently---attaining fast application completion such that users can promptly get the valid final output.
Nevertheless, compared with traditional workloads in OS and big data fields, LLM applications have two distinct characteristics. 
First, the resource demands of an LLM application (e.g., the token generation length of each request and the inter-request structure)---dependent to the runtime inputs---are uncertain a priori.
Given the difficulty to know the total application demand volume, existing serving systems like vLLM~\cite{vllm} and Parrot~\cite{lin2024parrot} choose a simple scheduling algorithm like FCFS, which hurts the scheduling efficiency due to the head-of-line blocking problem. 
Second, serving LLM applications often involves diverse backend resources (like the docker container for code testing~\cite{factool_code} or the KV cache for inference acceleration~\cite{touvron2023llama}), many of which are prepared in an on-demand manner~\cite{gao2024cost}. 
Consequently, the application completion time may be delayed due to the intermittent warm-up latency on cold backends. 
In summary, the absence of application demand information renders existing LLM serving systems inefficient in both queuing management and backend preparation.

The key to efficient serving of LLM applications is to obtain accurate demand information.
In fact, although LLM applications exhibit substantial demand dynamicity, it does not mean that the serving system has to be demand-agnostic: the application viewpoint brings promising opportunities for demand perception.
A typical LLM application is composed of multiple functional units (e.g., an inference task with a fixed system prompt to verify a just-generated claim); the resource demand of a given unit---due to its distinct functionality characteristics, is relatively stable across different runs. 
Since the LLM applications are usually recurring with their code files hosted on the cloud, it is possible to apply static and dynamic program analysis to model its resource demand.

However, making accurate demand modeling for general LLM applications is a non-trivial task. 
Different applications may involve different backend types and have different functional unit structures, and we need to design proper demand description primitives for generality.
Meanwhile, in each application run, the user input affects the triggered function units, and the demand volume on each function unit may also deviate from the average value previously profiled.
In that sense, any fixed demand representation would be over-assertive; meanwhile, we also need to conduct online estimation refinement to more precisely estimate the resource demands in the ongoing run.

In this paper, we design Hermes, an efficient system for serving LLM applications. 
In Hermes, we propose to model the resource demands of an LLM application as a \emph{Probabilistic Demand Graph} (PDGraph). 
A PDGraph organizes the diverse functional units of a LLM application with a graph structure: each PDGraph node describes the demand quantity of the corresponding functional unit with a distribution function, and records the downstream dependencies with an associated branch-taking probability.
%PDGraph supports online refinement of the estimated resource demands.
Moreover, by analyzing the demand correlation between upstream and downstream units, we can keep refining the demand estimation with the latest execution status of the ongoing run. % online estimation refinement to make more accurate estimation. 
With PDGraph, we can thus faithfully estimate, in a probabilistic manner, the resource demands of the entire application as well as of the upcoming unit.
%\chen{faithfully estimate the demand as a distribution function}
We can then leverage such information to optimize the queuing order and to determine the backend preparation moment.
%prepare appropriate backends in a timely manner.

In queuing optimization, the problem we now face is how to determine the scheduling order of applications whose demand is expressed as a distribution rather than a deterministic value.
In that case, scheduling applications under the classical shortest-remaining-time-first (SRTF) algorithm (based on the mean value of the demand distribution) is no longer optimal for minimizing the average appliation completion time. 
To adapt to demand uncertainty, Hermes adopts the Gittins policy.
% In optimizing the application queuing order, the problem we now face is to schedule jobs with unknown duration but known distribution, for which the classical shortest-remaining-time-first (SRTF) is not optimal. Hermes applies the Gittins policy based on the PDGraph models.
%A PDGraph describes an LLM application's overall demand as a distribution, and 
Gittins policy~\cite{scully2021gittins} has been proven optimal in scheduling jobs with \emph{unknown demands but known demand distributions}, a good fit for scheduling LLM applications. % whose overall demand is depicted as a distribution under PDGraph.
It works by calculating a Gittins index which is a runtime estimator of an application's true remaining processing time.
Additionally, we also consider the cases where each application is associated with a deadline; with the estimated demand distribution information, Hermes adopts the least-slack-time-first (LSTF) algorithm that prioritizes applications with higher risk of deadline violation.
% We also show how to leverage the PDGraph model for efficiency optimization in cases with user-specified application deadlines.  

Apart from queuing optimization, Hermes also leverages PDGraph for backend prewarming. 
With the demand information recorded in PDGraph, during the execution of a functional unit, we can prewarm the backends needed by its downstream units prior to their arrival. %to facilitate its fast execution.
Since a function unit in LLM applications may have multiple downstream units and their arrival times are also uncertain (depending on the execution time of the current unit), setting the backend type as well as the moment to prewarm presents as a clear trade-off. 
Given that prewarming wrong backends or prewarming backends too early would waste resources, Hermes introduces a knob to tune the trade-off between the latency reduction effect and the resource wastage incurred. 
Such a prewarming principle can be generally applied to both LLM backends (like KV cache) and non-LLM backends (like docker containers).

% Third, for the efficient execution of individual applications, we propose coinference-aware clairvoyant resource provisioning, applying the demand information from coinference modeling for management of the diverse resource backends.
% %  (all could be the service bottlenecks for LLM applications).
% %When managing the GPU memory storing KV cache~\cite{kwon2023efficient} and LoRA adaptors~\cite{hu2021lora}, we propose to determine the cache eviction order based on the coinference priority and the cache prefetching moments based on the profiled distribution of inter-stage gaps. 
% When managing the storage space of KV cache~\cite{kwon2023efficient} and LoRA adaptors~\cite{hu2021lora}, we propose clairvoyant cache management to determine the cache eviction order based on the coinference priority and the prefetching moments based on the profiled distribution of inter-stage gaps. 
% When managing non-LLM backends like CPU (to launch docker container) and network (to conduct RAG), we propose clairvoyant queueing that propagates the coninference priority to each resource backend.
% In cases with multiple LLM backends with heterogeneous batch size configurations, we propose clairvoyant backend selection, which means adaptively selecting the backend for each stage based on the profiled request parallelism. 
% These methods can ensure that an LLM application can have fast completion without being impeded on any resource backends.

We have implemented Hermes with over 4,000 lines of Python code, and have also prepared a workload suite containing representative LLM applications.
%, in two key modules---\texttt{AppHandler} and \texttt{HermesScheduler}.
The performance of Hermes is evaluated against mainstream serving systems with the diverse set of LLM applications. 
Experimental results show that Hermes can reduce the average application completion time by over 70\%, and can also make an improvement of over $1\times$ for cases with explicit deadlines. 
Moreover, ablation studies further confirm the effectiveness of Hermes in mitigating demand uncertainty as well as in efficient resource provisioning on diverse backends.

In summary, this paper makes the following contributions:
\begin{itemize}
    \item 	We identify the limitations of existing systems in serving LLM applications (which have distinct characteristics of dynamic demands and diverse backends), i.e., the long queuing delay and backend warm-up delay.
    \item 	We propose to model the demands of LLM application in a probabilistic and structured manner with PDGraph, and further design Hermes to leverage PDGraph for queuing order optimization as well as for backend preparation.
    \item 	We build a workload suite containing typical LLM applications, and confirm the effectiveness of Hermes with testbed experiments, demonstrating a salient efficiency improvement over existing systems.
\end{itemize}
%!TEX root = main.tex
\section{Background and Motivation}
\label{sec:background}

\subsection{LLM Applications: A Primer}
\label{sec:applications}

\begin{figure*}
    \centering
    \includegraphics[width=0.96\textwidth]{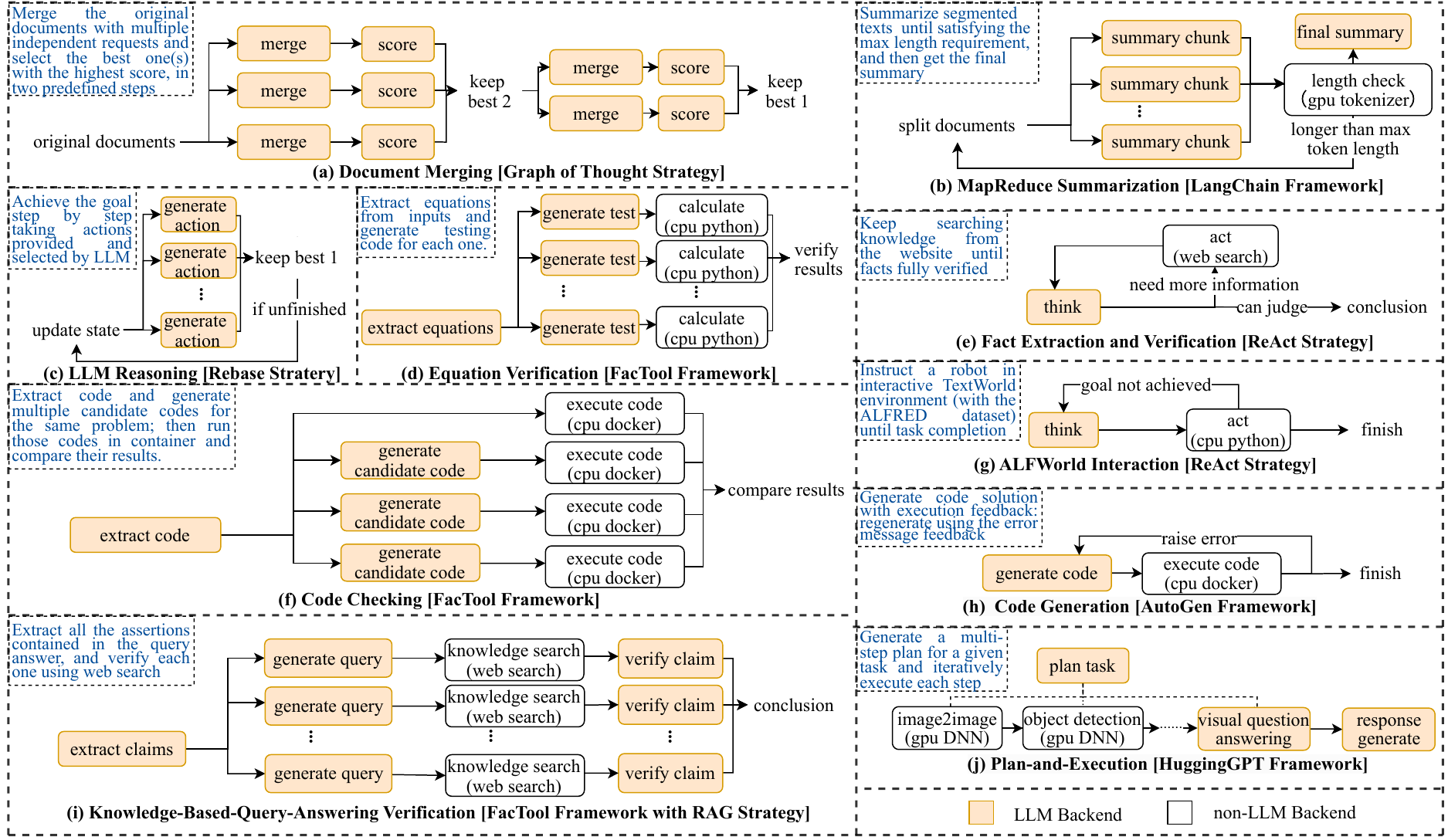}
    \caption{Ten representative LLM applications: 
    % For conciseness, hereafter we use the uppercase letters in each application name as the abbreviation representing that application. \chen{list each name and citation}
    (a) Document Merging (DM)~\cite{got_docmerge}; (b) MapReduce Summarization (MRS)~\cite{Langchain_MapReduce}; (c) LLM Reasoning (LLMR)~\cite{fu2024efficiently}; (d) Equation Verification (EV)~\cite{factool_math};  (e) Fact Extraction and Verification (FEV)~\cite{react_fever}; (f) Code Checking (CC)~\cite{factool_code}; (g) ALFWorld Interaction (ALFWI)~\cite{react_alfw}; (h) Code Generation (CG)~\cite{autogen}; (i) Knowledge-Based-Query-Answering Verification (KBQAV)~\cite{factool_kbqa}; (j) Plan-and-Execution (PE)~\cite{hugginggpt}.}
    %Used Applications. \chen{add brief introduction and source citation} \chen{add conversation and planning?} \chen{place description directly in figures? Summary the common patterns in title instead?} \gz{multi-turn conversation fault}}
    \label{fig:app_structure}
    \vspace{-0.1in}
\end{figure*}

\noindent\textbf{LLM applications.} 
Large language models (LLMs)~\cite{brown2020language,floridi2020gpt,touvron2023llama} with their built-in world knowledge can potentially be adopted for various fields like finance~\cite{li2023large}, arts~\cite{makridis2025impact} and science~\cite{ren2025towards}.
%Nonetheless, the capability of existing LLMs is limited, it is increasingly difficult to train new foundation models 
Due to the prohibitively high training cost and the exhausted data corpus sources,
% LLMs with subt
%can potentially revolutionize various domains~\cite{dosovitskiy2020image,hadi2023survey}. 
%Yet, even though LLMs are evolving rapidly~\cite{sonoda2024diagnostic,reid2024gemini}, it is commonly recognized that a single LLM request is often deficient for realistic problems. 
%Given the prohibitively high cost to train newer-generation LLMs, 
there is an emerging trend to boost the power of existing LLMs by \emph{scaling-out} instead of by \emph{scaling-up}~\cite{compound_ai}. 
%In a word, 
%To be specific, by combining a set of LLM inference tasks and as well as the necessary non-LLM tasks (like knowledge retrieval and tool calling), users can improve the LLM performance in aspects like input length~\cite{jin2024comprehensive,lin2024infinite}, output reliability~\cite{chern2023factool,ji2023towards} and functional flexibility~\cite{shen2024hugginggpt,shen2024llm}.
%To summarize, 
That is, to enlarge the input length~\cite{jin2024comprehensive,lin2024infinite}, improve the output reliability~\cite{chern2023factool,ji2023towards}, or integrate more functionalities~\cite{shen2024hugginggpt,shen2024llm}, LLM practitioners often need to jointly execute multiple LLM inference tasks and the necessary non-LLM tasks, which form a compound task workflow---we call an \emph{LLM application}.
To elaborate, we create a comprehensive workload suite of LLM applications, as shown in Fig.~\ref{fig:app_structure}.
That workload suite involves ten representative applications built with specific frameworks (e.g., FacTool~\cite{chern2023factool}) or following certain control strategies (e.g., Graph of Thought~\cite{besta2024graph} and ReAct~\cite{ji2023towards}).

In particular, we note that LLM applications differ from traditional workloads (in OS or big data) in \emph{demand dynamicity} and \emph{backend diversity}.
First, the execution status of an LLM application is highly dynamic: due to mechanism like ReActing~\cite{ji2023towards,yao2022react} and LLM-planning~\cite{shen2024hugginggpt}, the inter-task structure can only be determined at runtime; meanwhile, both the input and output token lengths of each LLM task are also unknown beforehand~\cite{gao2024cost,wu2023fast}. 
Second, serving LLM applications requires provisioning diverse LLM and non-LLM backends.
Regarding LLM backends, different applications may prefer different foundation models or fine-tuned adapters~\cite{hu2021lora,chen2023frugalgpt}.
Regarding non-LLM backends, a diverse set of backends may also be needed: for example, in the code-generation application, the LLM-generated codes need to be tested on CPU backends (e.g., with a docker container~\cite{factool_code,docker_execution} to provide an isolated environment); meanwhile, the HuggingGPT application~\cite{hugginggpt} may require loading diverse non-LLM models 
%(e.g., Stable Diffusion~\cite{rombach2022high}) 
as tools.

%To summarize, to enlarge the input length, improve the output reliability, and integrate more functionalities, LLM practitioners often need to conduct multiple inferences with the necessary non-LLM computations, which form a compound workflow.
% \ys{What exactly are LLM applications? Is LLM-based chat an LLM app? Yes, it is. I think the main trend is that to mitigate LLM hallucination and to improve AI accuracy at complex reasoning tasks, LLM-based apps have evolved from single-turn LLM-based chat to a more compound and complicated workflow. Then list examples such as RAG, o1, factool etc.}

% In summary, to address real-world tasks—such as handling large input sizes, ensuring output reliability, and supporting diverse capabilities—LLM practitioners often perform multiple inferences alongside necessary non-LLM operations, collectively referred to as an \emph{LLM application}. % compound workflow.
% %We call the above compound workflow---with multiple LLM requests and even non-LLM tasks to accomplish a realistic mission---as an \emph{LLM application}.
% LLM applications are rapidly emerging in recent years~\cite{hadi2023survey,tan2024teola,lin2024parrot}, and would become a mainstream workload type in the GenAI era.

\vspace{0.3em}
\noindent\textbf{Cloud-based serving of LLM applications.} 
%\chen{add a architecture figure to show the queueing + dispatching modules}
LLM applications, just like OpenAI assistant API~\cite{assistants_api}, are usually hosted on the cloud for ease of user access~\cite{multi_agent,ibne2024ensuring,sheng2024fairness,lin2024parrot}.
The code files of those applications are maintained by the service provider, which would be launched when users submit their application inputs.
When serving multiple LLM applications from different users~\cite{sheng2024fairness,ibne2024ensuring}, the service provider needs to first organize the incoming tasks in a global queue, and then respectively dispatch each task to the proper LLM or non-LLM backend.
%Such scheduling (i.e., \emph{queuing} and \emph{dispatching}) policies crucially affect the user experience and service profit, and 
A good serving system shall attain high execution efficiency and also high resource utilization.
%For the service provider, there are two scheduling procedures: 
%When executing an LLM application, both LLM requests and non-LLM operations must be dispatched to appropriate resource backends, such as GPU servers or---for code testing in a sandbox environment---Docker servers.
%A production service provider concurrently serves many LLM applications from different users~\cite{sheng2024fairness,ibne2024ensuring}; 
Specifically, when launching an LLM application, users primarily focus on the \emph{application completion time}---that is, the time when the high-quality output (like the valid code) is finally returned~\cite{lin2024parrot}. %\chen{add a footnote to explain why not include TTPT} 
%For good user experience and high service profit, 
Therefore, the efficiency objective shall be set at the application level, i.e., minimize the average application completion time (ACT) or---if the expected deadline is provided---maximize the goodput (number of applications that complete before the deadline). 
%It is thus desirable to attain efficient scheduling at the application level, i.e., the serving system needs to minimize the average application completion time (ACT) or---if the expected deadline is provided---maximize the goodput (number of applications that complete before the deadline). 
Meanwhile, for high resource utilization, LLM service provider shall avoid holding resources idly in serving the LLM applications.

%\chen{add more explanations on scheduling objectives}

\subsection{Existing Practices and Their Limitations}
\label{subsec:limitation}

For LLM applications, due to their strong demand dynamicity and backend diversity, existing serving systems often fail to yield high scheduling efficiency. % provide efficient serving performance.

First, confronting uncertain resource demands, existing practices usually adopt a simple heuristic in task queuing (like FCFS), which is known to be inefficient.
A mainstream LLM serving framework nowadays is vLLM~\cite{vllm}, which schedules LLM inference tasks in a FCFS manner---without awareness to the high-level existence of LLM applications; in resource contention, the constituting tasks of an LLM application may be interleaved by tasks from other applications, suffering delayed application completion. 
%compromising the end-to-end performance.
Recently, Parrot~\cite{lin2024parrot} proposes to schedule the inference tasks of an application together; however, without concrete knowledge of resource demand volume, it schedules LLM applications also following FCFS, thus suffering the head-of-line-blocking problem.
Our testbed measurement shows that, executing a small application (KBQAV in Fig.~\ref{fig:app_structure}(i)) after a large application (DM in Fig.~\ref{fig:app_structure}(a)) yields an average ACT of 52.8s, yet with the reversed order it is only 40.3s. % from 40.3s to 52.8s compared to the reversed order.  
Another work VTC~\cite{sheng2024fairness} proposes to fairly allocate the computing capability between different applications (tenants) to avoid head-of-line blocking, yet it still suffers low efficiency by forcing each application to use only its fair share under contention.

\begin{figure}[t]
    \centering
    \subfloat[Cold LLM Backends]
    {
        \centering
        \includegraphics[width=0.45\linewidth]{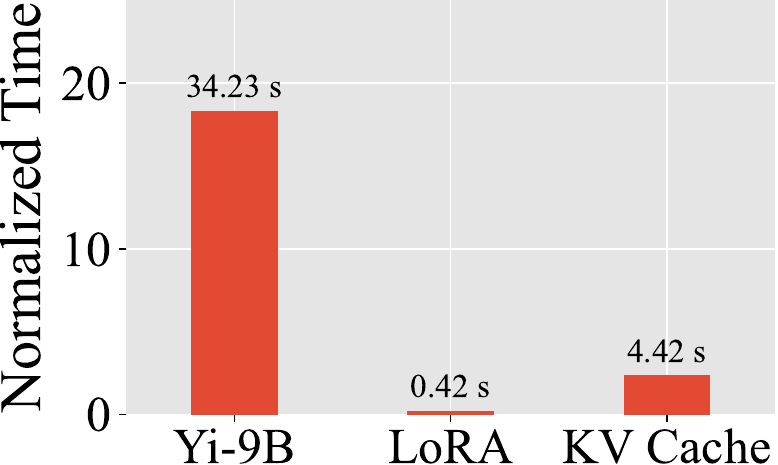}
        \label{fig:cold_llm}
    }
    \hfill
    \subfloat[Cold non-LLM Backends]
    {
        \centering
        \includegraphics[width=0.45\linewidth]{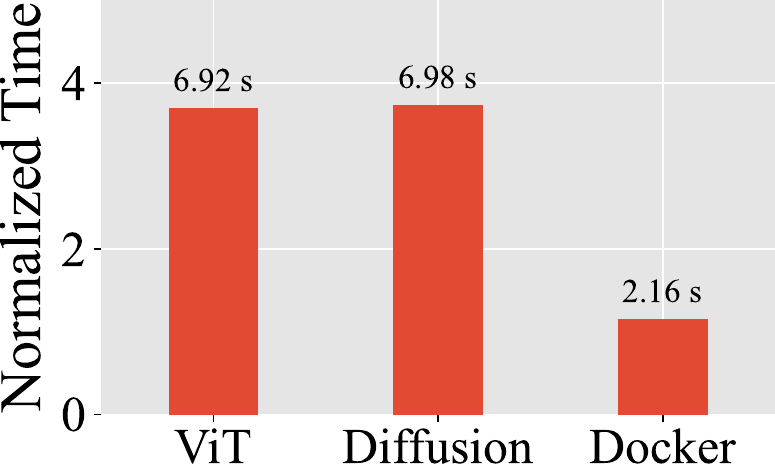}
        \label{fig:cold_nonllm}
    }
    \caption{Costs to warm up code backends (normalized to the execution time of a typical LLM inference with 1000/100 input/output tokens). The LoRA is for LLaMA-7B with rank 8 and the KV cache size is 128K (loaded to A100 GPUs). The docker image is \texttt{python:3.10-slim}.}% used are two V100 of cold Backends. \chen{add model/lora/kv-cache/docker information} (a) The time we load Yi-9B, LoRA (rank=8) from disk to GPU and KV Cache of 128k token of from CPU to GPU on 2 V100.}
\end{figure}

Second, since LLM applications may call diverse backends at uncertain moments, it is inefficient to make such backends standby all the time.
For example, among the 130 official input examples provided by HuggingGPT~\cite{shen2024hugginggpt}, our measurement shows that only 20 runs triggered the text-to-video model. %\chen{can be extended to a table}
To avoid resource wastage, such serving backends shall be prepared in an on-demand manner, which however incurs cold start delay on diverse backends. 
%it is hard to timely prepare the desired backends, which would incur either low resource utilization or cold start delay. %many backends need provisioned in a serverless manner to avoid resource wastage, yet this would incur substantial cold start delay.
%As elaborated in \Cref{sec:applications}, LLM applications may potentially demand a wide array of diverse backends, it is inefficient to make such backends standby all the time.
For instances, regarding LLM backends, users may choose different foundation models based on their cost-quality preference~\cite{chen2023frugalgpt} or require customized LoRA adaptors~\cite{hu2021lora} for domain-specific capability, and meanwhile KV-cache reuse may also be adopted to accelerate inference speed~\cite{kwon2023efficient,gao2024cost}; such model or cache content need to be pre-loaded into the GPU HBM before inference starts.
Regarding non-LLM backends, for code-generation applications the docker containers need to be launched before code testing, and for HuggingGPT applications~\cite{shen2024hugginggpt} the DNN models like object recognition also need to be loaded into GPUs before task execution.  
Those preparation processes may take non-negligible time and slow down the end-to-end application completion.
%(even user-defined ones like OpenAI function calling~\cite{function_calling}), and it is inefficient to make such backends standby all the time. 
%For example, among the 130 official input examples provided by HuggingGPT, our measurement shows that only 20 runs triggered the text-to-video model. 
%To avoid resource wastage, those infrequent model backends shall be provisioned in a serverless manner---launched after demand arrival.
%Moreover, for LLM backends, 
%in cases with cross-request token reuse, the KV-cache~\cite{kwon2023efficient,gao2024cost} of must also be loaded in advance for fast .
    %Therefore, the backend may be cold at request time, requiring additional time for preparation.
    % For example, to prepare the LLM backends, we need to load the LLM model into the GPU HBM, and in some cases with token reusing or model customization, the KV-cache and LoRA adapters also need to be loaded.
    %For instance, preparing LLM backends involves loading the LLM model into GPU HBM. 
   % Moreover, in cases with token reuse or model customization, the KV-cache~\cite{kwon2023efficient} or LoRA adapters~\cite{hu2021lora} must also be loaded before inference starts.
    %For non-LLM backends, it takes time to warm up the desired docker container, and the DNN models for multi-modal recognition or generation also need to be loaded into a GPU.
    In Fig.~\ref{fig:cold_llm} and Fig.~\ref{fig:cold_nonllm}, we show the typical warm-up time of a series of backend contents, which are up to  $18\times$ of the inference time of a typical inference task (with 1000 input tokens and 100 output tokens). 
    %In particular, 
    Noticing the need to conduct KV cache prewarming, the recent CachedAttention work~\cite{gao2024cost} proposes to conduct KV cache eviction and prefetching based on the scheduler waiting queue, a method we call Evict/Prefetch-Waiting-Queue (EPWQ). 
    %However, EPWQ treat the KV cache desired by a task not yet submitted at the lowest priority; when a down, its KV cache may be swapped out, downstream task  which would be easily swapped out and triggers cold start delay when the downstream task arrives.
    However, EPWQ cannot prefetch the KV cache for unbacklogged requests (e.g., a future request not yet submitted); 
    %without the application view, after an inference task in a high-priority application completes, 
    because EPWQ does not have the application viewpoint, a downstream task of an LLM application---even if it has the highest priority---may still bear a KV cache loading delay after arrival.
    %has the highest priority, the  downstream request cannot 
%LRU conducts reactive swapping with prewarming, and EPWQ prewarms the KV cache only when the request is already in the waiting queue---which is often too late.
%means to determine the priority of a cached content based on its request order in the local waiting queue;
% conduct cache eviction based on the pending request queue (adopted in CachedAttention~\cite{gao2024cost}); 
%yet, during the inter-stage gap of an application, since the downstream request is not in the queue, its desired KV cache would be of the lowest priority
    %It shows that such warm-up overheads are indeed substantial (up to $18\times$). %, and must be addressed for fast serving of LLM applications. 
%\chen{add some attempts to address those issues? Serverless LLM for model loading and cachedattention for KV loading?}

To summarize, confronting dynamic demands on diverse backends, existing LLM serving systems have to employ static queuing policies and also conduct reactive resource provisioning.
Such a \emph{demand-agnostic} serving methodology hurts the overall efficiency of LLM applications.

\subsection{Insight and Challenges}
\label{sec:modeling}

\begin{figure*}
		\centering
		{
			\centering
			\includegraphics[width=3.2cm]{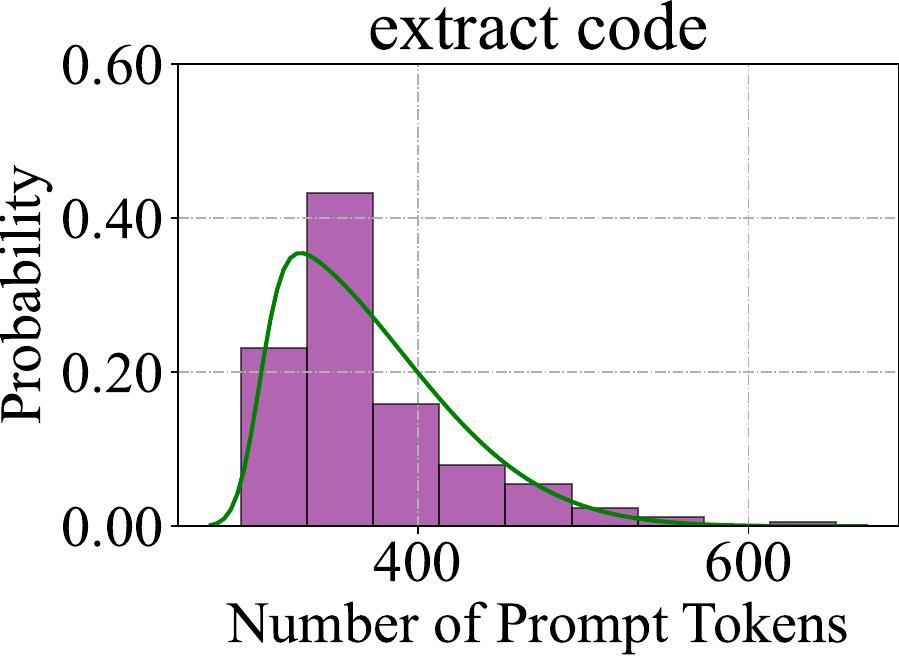}
			\label{fig:p_token_factoolcode}
		}\hfill
		{	
			\centering
			\includegraphics[width=3.2cm]{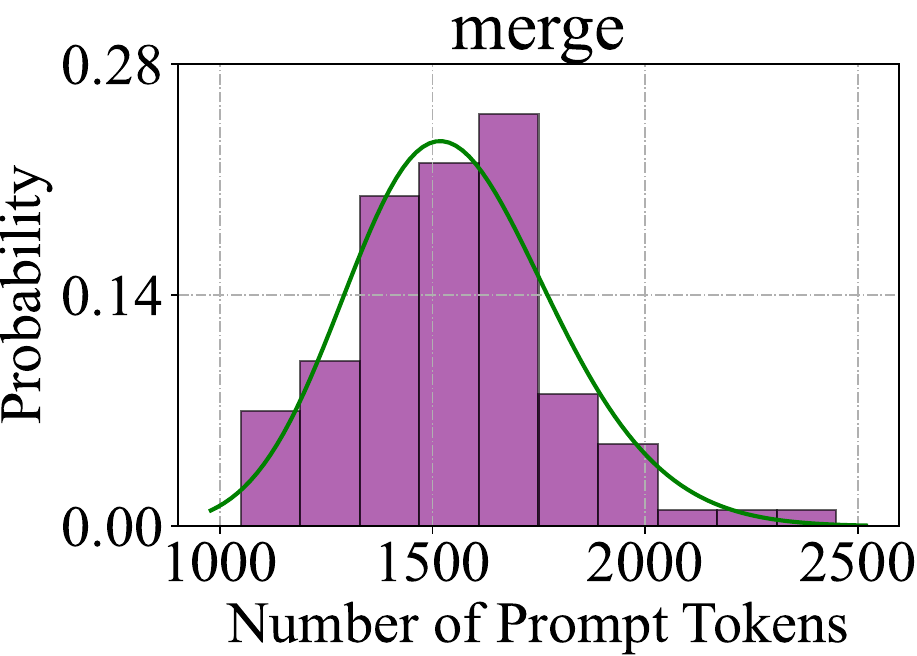}
			\label{fig:p_token_gotdecmerge}
		}\hfill
		{	
			\centering
			\includegraphics[width=3.2cm]{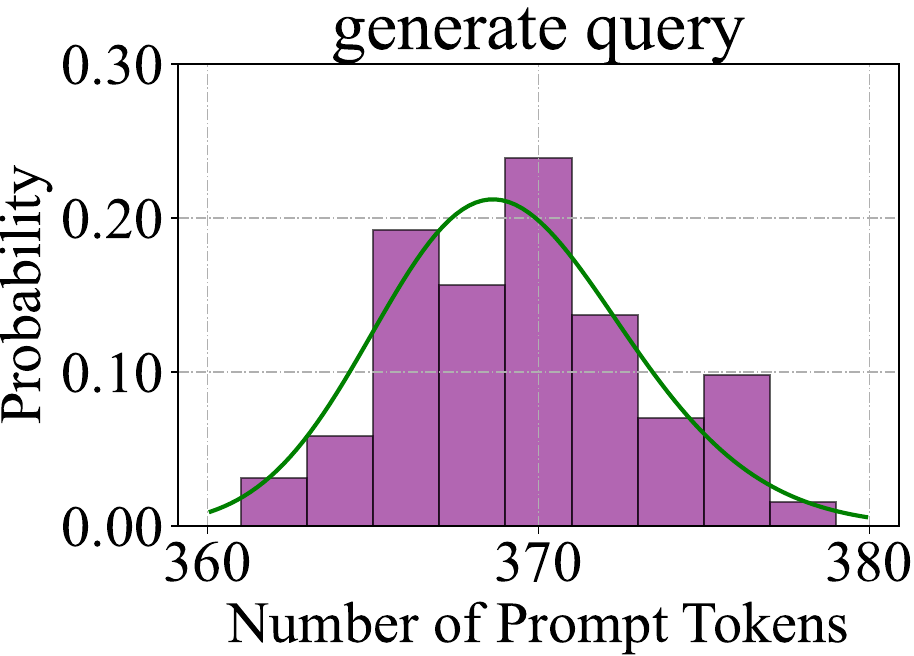}
			\label{fig:p_token_factoolkbqa}
		}\hfill
		{	
			\centering
			\includegraphics[width=3.2cm]{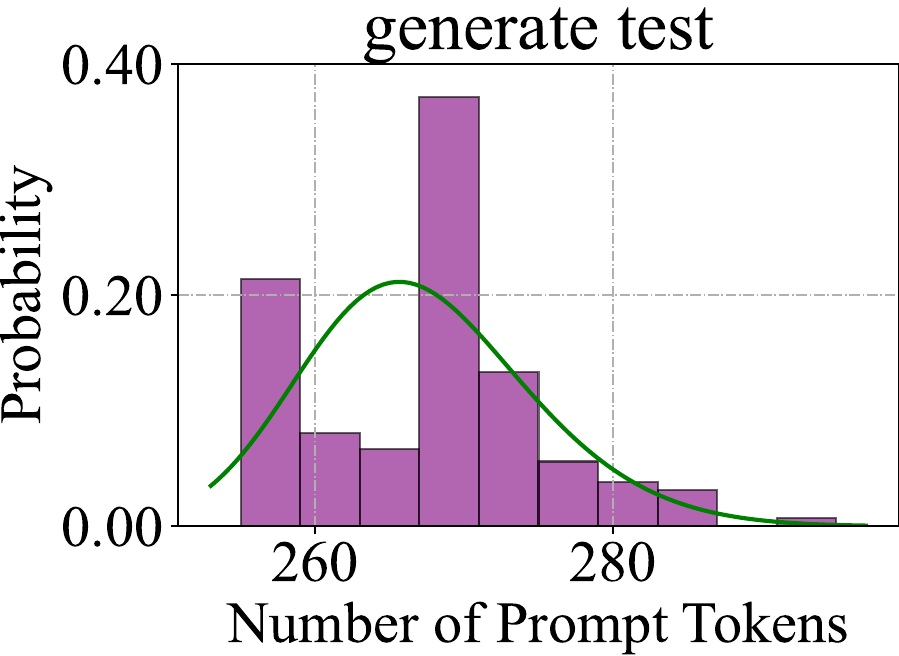}
			\label{fig:p_token_factoolmath}
		}\hfill
		{	
			\centering
			\includegraphics[width=3.2cm]{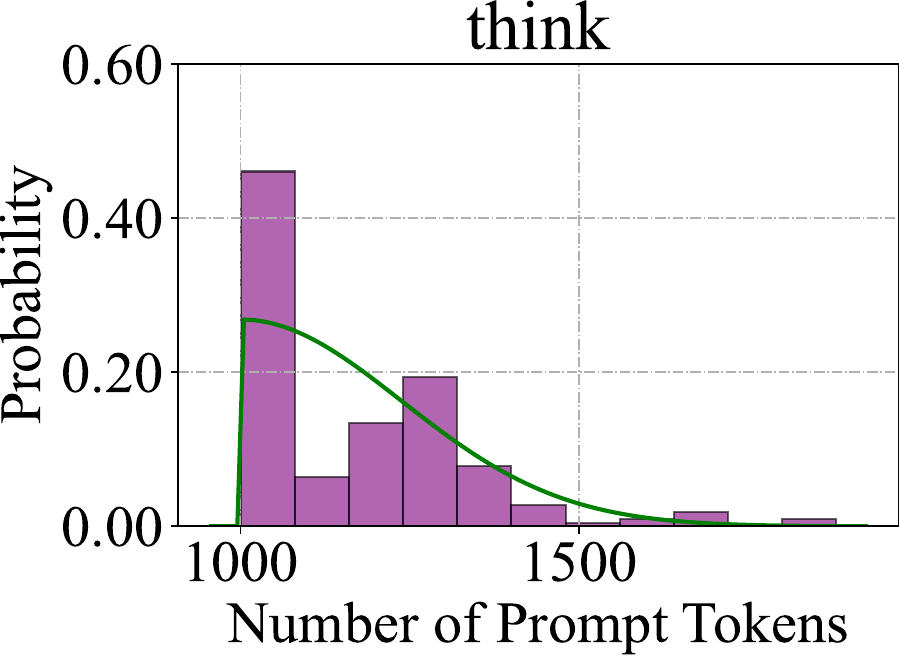}
			\label{fig:p_token_reactfever}
		}\hfill
     
		{
			\centering
			\includegraphics[width=3.2cm]{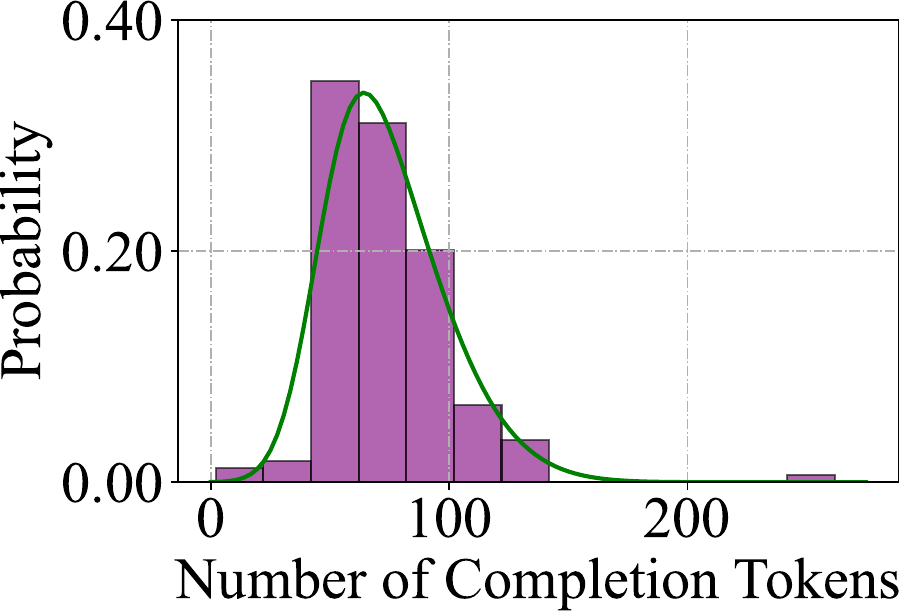}
			\label{fig:c_token_factoolcode}
		}\hfill
		{	
			\centering
			\includegraphics[width=3.2cm]{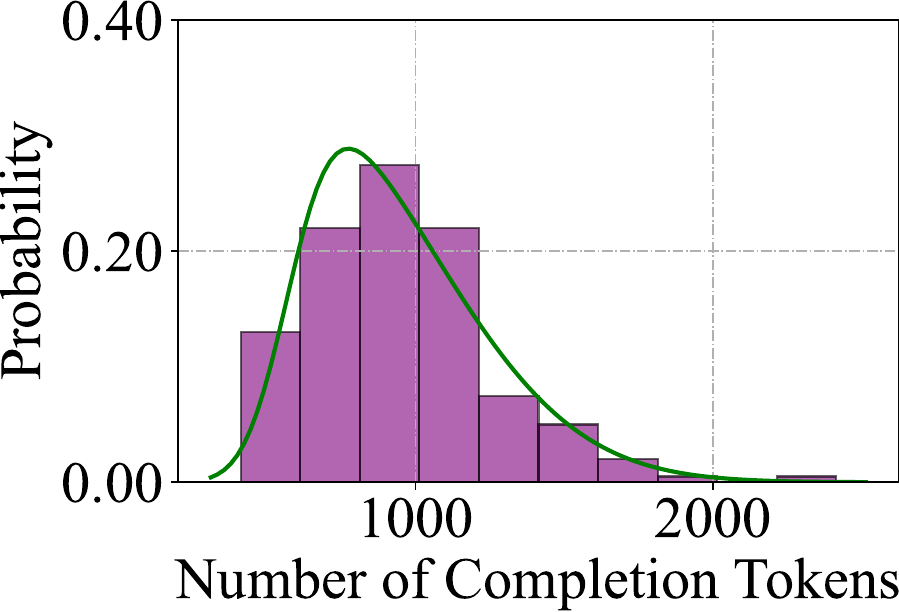}
			\label{fig:c_token_gotdecmerge}
		}\hfill
		{	
			\centering
			\includegraphics[width=3.2cm]{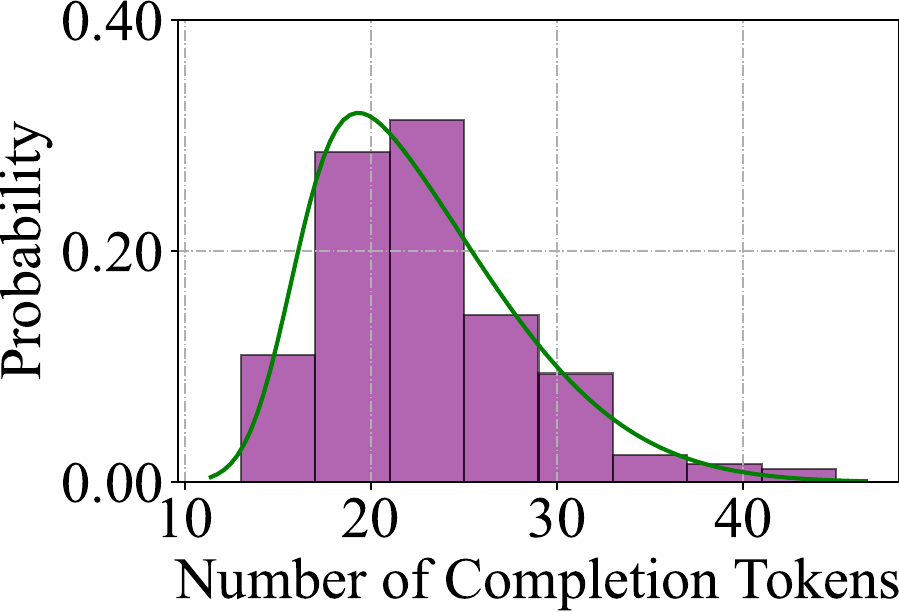}
			\label{fig:c_token_factoolkbqa}
		}\hfill
		{	
			\centering
			\includegraphics[width=3.2cm]{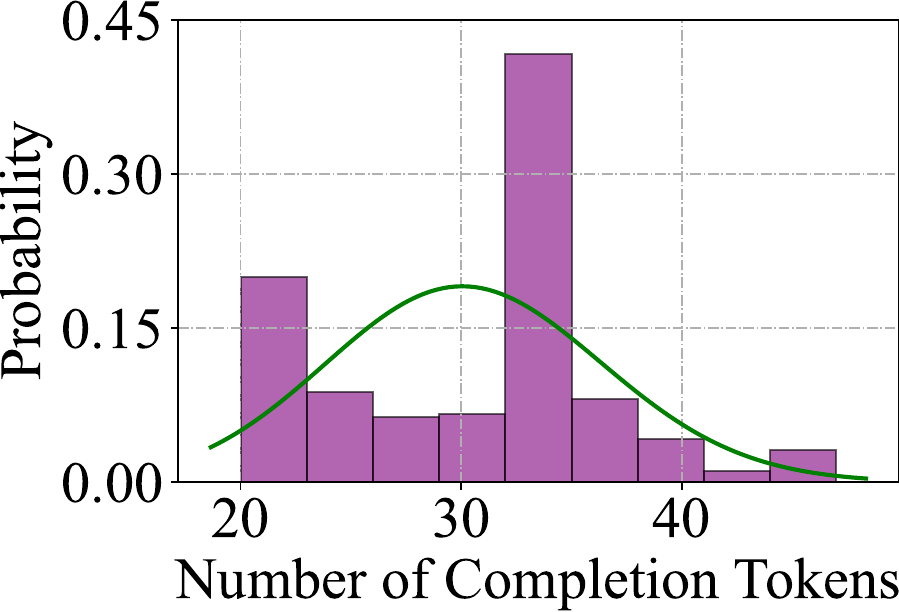}
			\label{fig:c_token_factoolmath}
		}\hfill
		{	
			\centering
			\includegraphics[width=3.2cm]{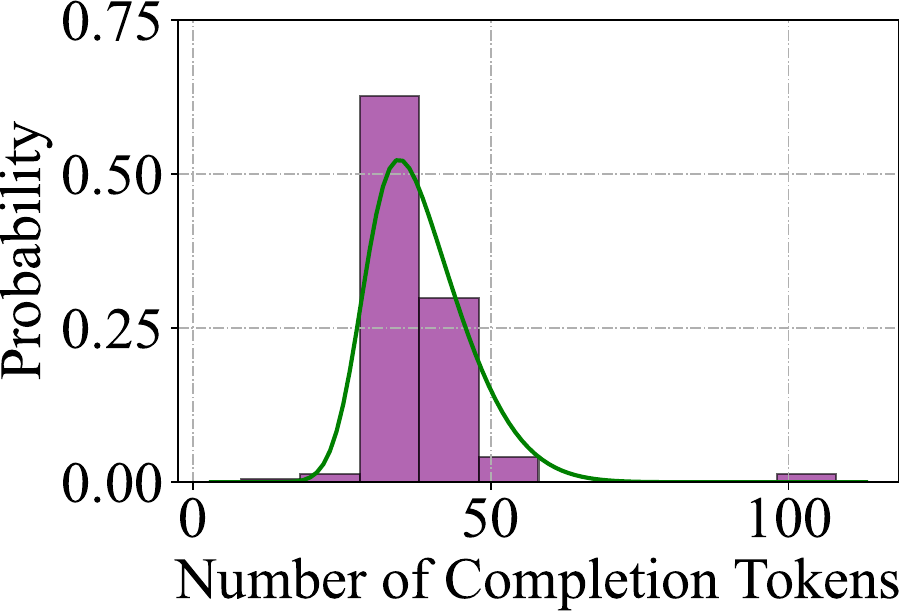}
			\label{fig:c_token_reactfever}
		}\hfill
        \vspace{-.1in}
        \subfloat[Code Checking]
		{
			\centering
			\includegraphics[width=3.2cm]{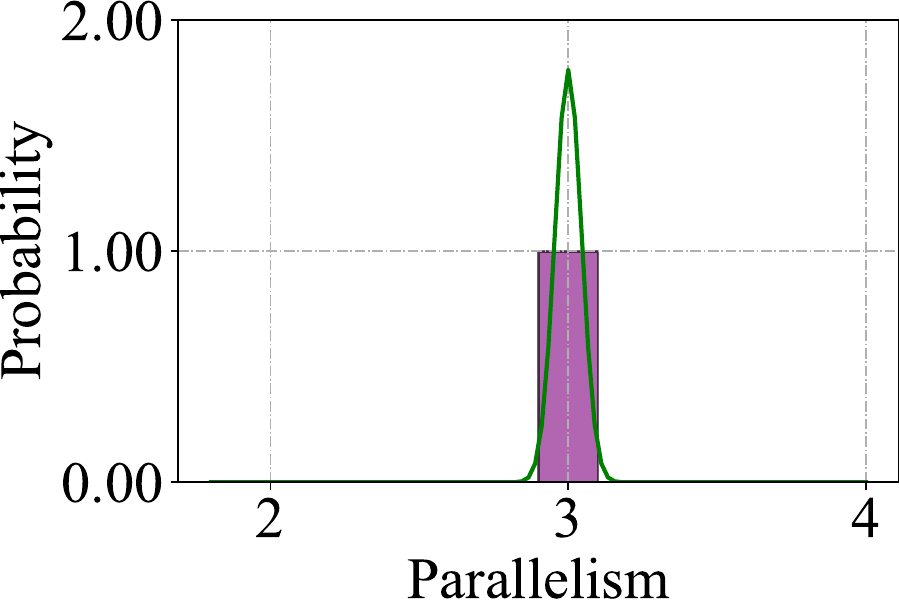}
			\label{fig:p_factoolcode}
		}\hfill
		\subfloat[Document Merging]
		{	
			\centering
			\includegraphics[width=3.2cm]{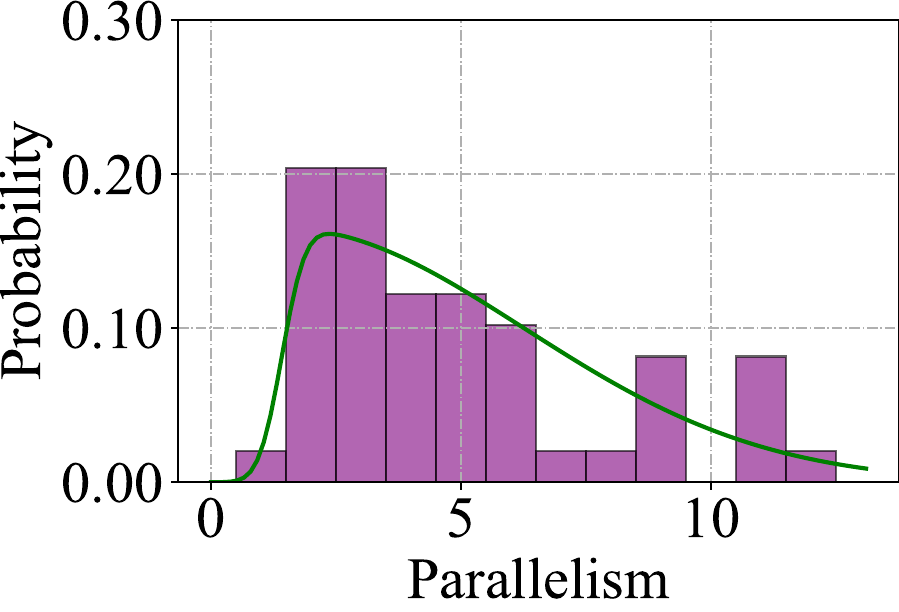}
			\label{fig:p_mapreduce}
		}\hfill
		\subfloat[KBQAV]
		{	
			\centering
			\includegraphics[width=3.2cm]{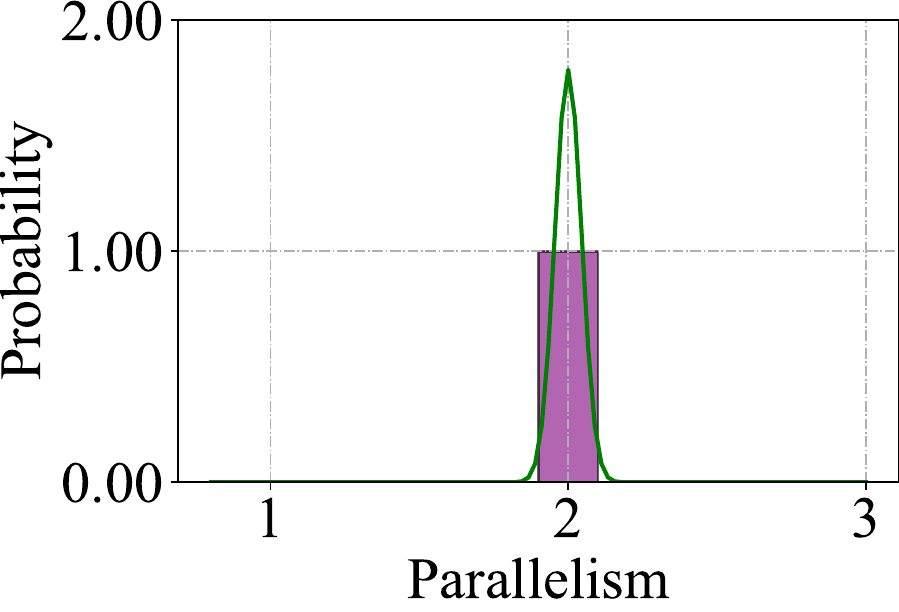}
			\label{fig:p_factoolkbqa}
		}\hfill
		\subfloat[Equation Verification]
		{	
			\centering
			\includegraphics[width=3.2cm]{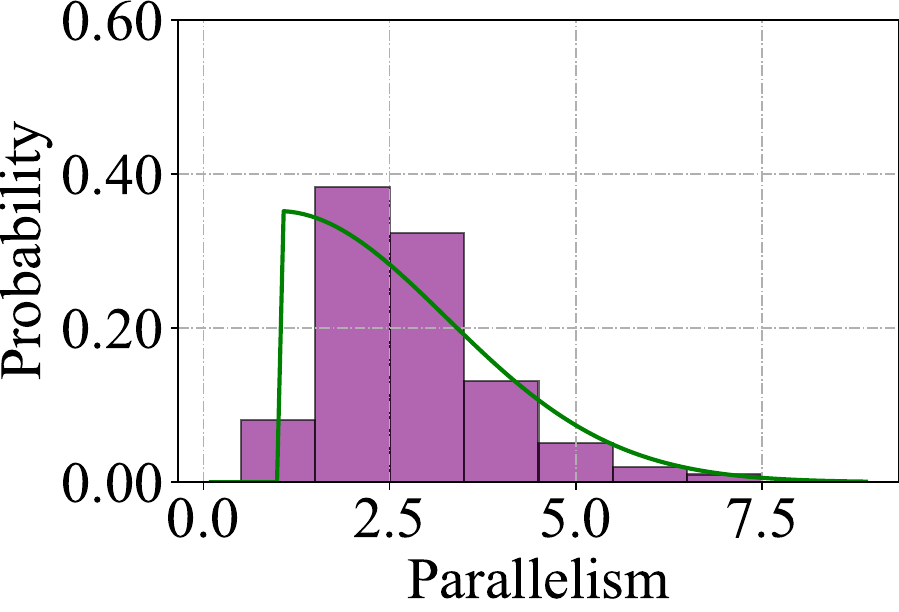}
			\label{fig:p_factoolmath}
		}\hfill
		\subfloat[FEV]
		{	
			\centering
			\includegraphics[width=3.2cm]{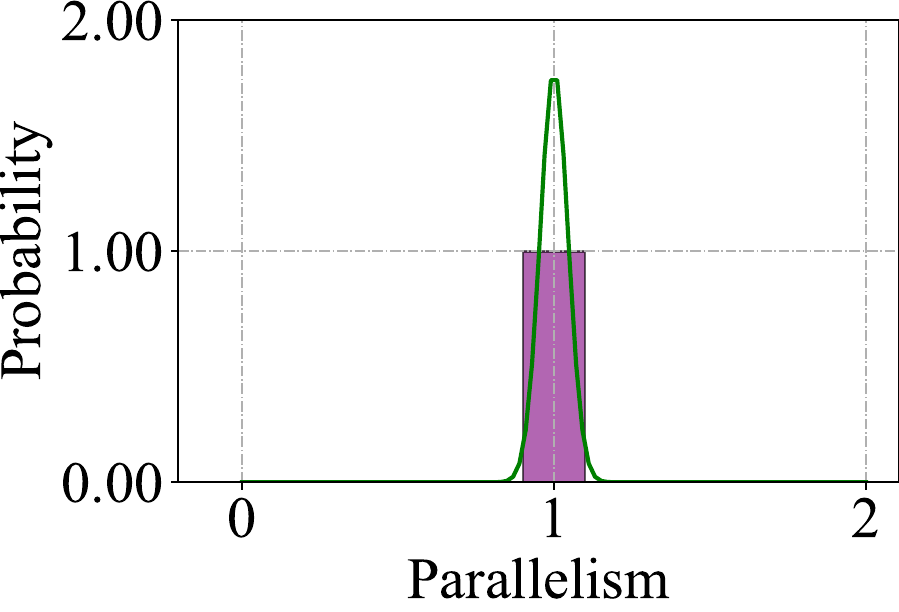}
			\label{fig:p_reactfever}
		}
            \vspace{-.1in}
		\caption{
                Prompt and completion token length distribution for an arbitrarily-selected request (marked at the top)---in five applications each conducting over 100 trial runs.
                In each case, we divide the length range into 10 buckets, and calculate the value appearance probability in each bucket (accompanied by the fitted curves assuming skewed Guassian distribution for reference).%\chen{add parallelism predictability}}
		\label{fig:token_usage}
	}
    \vspace{-0.1in}
\end{figure*}

\phm{Insight.} 
To enhance the serving efficiency of LLM applications, the key is to obtain their demand information in advance. 
While the demands of LLM applications are dynamic, we note this does not mean that they are totally unpredictable.
%such that LLM applications must be served in the dark.
%On the one hand, efficiency enhancement does not require the exact demand information~\cite{fu2024efficient} (an approximate estimation may suffice); on the other hand, 
In fact, LLM applications are usually recurring, making it possible to make thorough performance profiling offline.
As shown in Fig.~\ref{fig:app_structure}, each LLM application is composed of multiple functional units: tasks in an LLM unit (e.g., the \texttt{merge} unit in the DM application) share the same system prompt, and tasks in a non-LLM unit require distinct backend type (e.g., the \texttt{execute-code} unit require specific docker container in CG application).
With program analysis, we can profile the execution information of each function unit, which can potentially be applied for demand estimation.
%a non-LLM function 
%in \Cref{sec:background}, an LLM application is composed of a series of LLM and non-LLM tasks. 

To elaborate, we execute five LLM applications from Fig.~\ref{fig:app_structure} for 100 times each, replaying different inputs sampled from the official dataset.
Fig.~\ref{fig:token_usage} presents the execution statistics, including input/output token lengths and request parallelism, for an arbitrarily selected functional unit of each application.
%In particular, we divide the length range for each request position into 10 buckets, and calculate the probability of each bucket (accompanied by the fitted curves assuming skewed Guassian distribution, for reference). 
As illustrated in Fig.~\ref{fig:token_usage}, while the resource demands are non-deterministic, they still exhibit significant stability across multiple trial runs.
%\chen{For example, the input token length of the Document Merging application is always larger than 1000, yet for KBQAV application it is always between 360 and 380. }
For example, the output token length of the \texttt{merge} request in DM application is around 1000, yet for the \texttt{generate-query} request in KBQAV application it is between 10 and 50.
That phenomenon is indeed reasonable because demand volume is to some extent an inner property based on the usage scenario of an LLM application:
the output of the \texttt{merge} request is a document, whereas the \texttt{generate-query} request produces a short query. Consequently, the latter's length distribution exhibits a significantly smaller mean and deviation.
Such stability also exists in the request parallelism aspect. 

Therefore, we can potentially exploit dynamic program analysis to estimate the resource demands of LLM applications.
Such information can help optimize task queuing and backend provisioning for the LLM serving system.
Ideally, if we know the total execution time of each LLM application, we can apply shortest-remaining-processing-time (SRPT) to minimize the average ACT; if we know which backend to use next and the moment to use it, we can calculate the best time to prewarm the cold backend, attaining fast application completion without any resource wastage. 
% the historical statistics can help to narrow down the estimation range and facilitate accurate prediction. 
%As mentioned in \Cref{sec:background}, an LLM application is written as a fixed code segment hosted on the cloud; such a code segment describes the control flow of different \emph{functional units} composing that LLM application: some units issue requests (with the same system prompt) to the LLM backends,
%(each unit has a distinct system prompt shared by the requests it issues), 
%yet some other units launch non-LLM operations. 
%Based on such a structured view of functional units, we can leverage their correlations to estimate the demands of each LLM application as accurately as possible.
%To be specific, we find that such demand correlation exists in two aspects.
%First, for each functional unit, its resource demands across different trial runs are often correlated.
%Second, in the ongoing run of an LLM application, the resource demands of dependent functional units are also correlated.
%Next, we elaborate on them with empirical measurements. %\chen{mention prediction works?}

\phm{Challenges.}
Yet, while promising, it is however a non-trivial task to leverage program profiling of LLM applications for the best efficiency performance, and the challenges are twofold.

On the one hand, demand dynamicity and backend diversity are built-in properties of LLM applications, and the demand modeling must be conducted in a general and also accurate manner.  
By ``general'', the modeling method needs to be compatible with diverse backends and variant demand dependency structures; by ``accurate'', the modeling method needs to maintain a proper level of uncertainty---trying to narrow down the estimation range yet without being over-assertive. 
To elaborate, in each application run, the type/number of functional units and the task durations---which crucially relate to the runtime inputs---may substantially deviate from the average case of the profiling results; simply recording the average profiling results would be inaccurate.
% We therefore must preserve the variation information in demand modeling, and keep refining the demand estimation for each ongoing run by exploiting the runtime execution information

On the other hand, profiling-based performance modeling can only mitigate but not eliminate demand uncertainty, rendering it hard to directly attain the optimal serving performance.
In fact, we will show later that the classical SRTF policy is no longer optimal when scheduling jobs with uncertain demand. 
Therefore, our queue management and backend prewarming methods must adapt to such uncertainty for the best possible performance. 
We will address such challenges in the later section.
\section{Hermes Design}
\label{sec:solution}

\subsection{Overview}

\begin{figure}
    \centering
    \includegraphics[width=0.9\linewidth]{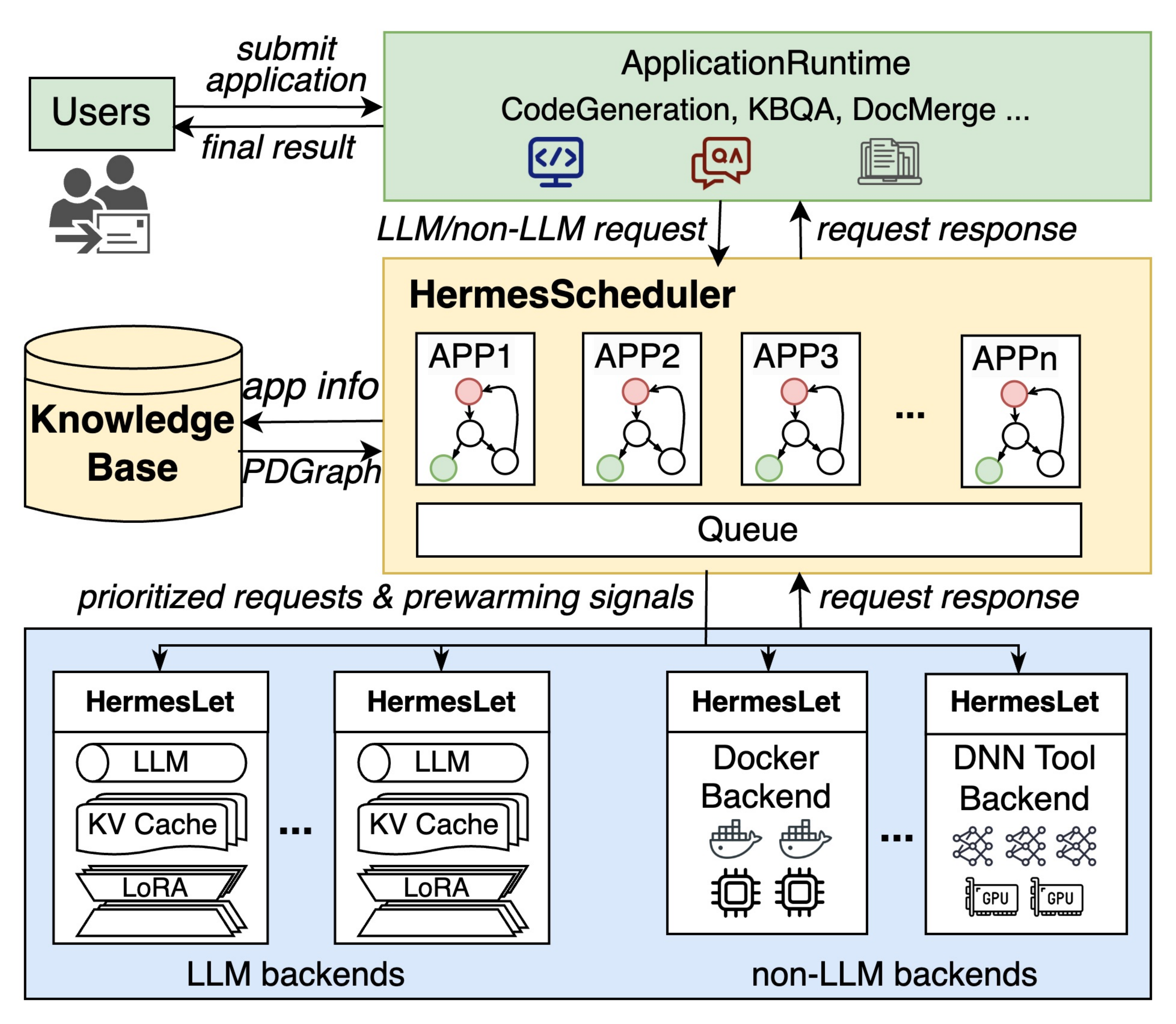}
    \caption{The system architecture of Hermes.}
    \label{fig:overview}
\end{figure}

We present Hermes, an efficient serving system for LLM applications hosted on clouds.
Hermes is built upon a performance modeling paradigm called \emph{Probabilistic Demand Graph} (PDGraph), which can yield accurate demand estimation for general LLM applications in an \emph{offline+online} manner (\cref{subsec:demand_modeling}).
As shown in Fig.~\ref{fig:overview}, Hermes maintains a knowledge base storing the profiled information for each application.
Once a user launches a cloud-hosted LLM application with its tasks submitted to the \texttt{HermesScheduler}, the \texttt{HermesScheduler} retrieves the PDGraph for that application to assist the scheduling process.
To be specific, the \texttt{HermesScheduler} schedules each task based on the application priority it belongs to, which is calculated based on the estimated demand information (\cref{subsec:scheduling}). 
Moreover, based on the instantaneous execution status, the \texttt{HermesScheduler} also issues backend prewarming signals to the \texttt{HermesLet} at proper moments to avoid code start latency (\cref{subsec:provisioning}).
	
\subsection{Demand Modeling with PDGraph}
\label{subsec:demand_modeling}

In this part, we seek to conduct accurate demand modeling for general LLM applications, with the objective to optimize queue management and backend prewarming strategies.
Recall that in Fig.~\ref{fig:token_usage}, we learned that the demand volume of a given functional unit is relatively stable across different trial runs.
However, the inter-unit structure---due to mechanisms like reacting and LLM-planning---is not determined apriori, and meanwhile the specific demand volume in a single run may deviate from the average case. 
Therefore, in demand modeling of LLM applications, we need to cover the dynamicity of inter-unit structure for \emph{generality}, and also combine historical execution information with runtime hints for \emph{accuracy}. 
To that end, we propose to model the LLM applications as a probabilistic graph of functional units, and assign proper properties to each unit node.
We call thus a modeling paradigm as \emph{Probabilistic Demand Graph} (PDGraph).

% We employ offline profiling to construct a knowledge base containing all LLM applications hosted on the cloud.
% To model the dynamic and structured resource demands of an LLM application, we propose \emph{Probabilistic Demand Graph} (PDGraph).
% Our objective here is to design a (1) \emph{general} representation method covering diverse LLM applications and, in the meantime, (2) as \emph{precise} as possible. 

\begin{figure}
    \centering
    \includegraphics[width=0.7\linewidth]{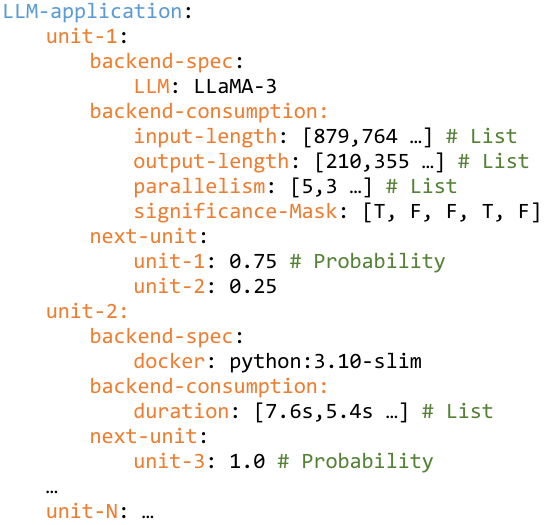}
    \caption{PDGraph example of an LLM application.}% Each functional unit only has one backend not set to None.\chen{fewer backends}} 
    \label{fig:demand_model}
\end{figure}

\phm{Demand modeling for general LLM applications with PDGraph.}
As shown in Fig.~\ref{fig:demand_model}, in our demand modeling paradigm with PDGraph, each application is recorded as a list of functional units. 
For each functional unit, we record three data types: \emph{backend-spec}---describing the resource type and configuration specifics, \emph{backend-consumption}---describing resource consumption amount on that backend, and \emph{next-unit}---describing the probabilistic jumping relationship between dependent functional units.
In this way, we can model the resource demands of general applications with diverse structures and backends: each backend (even user-defined ones like OpenAI function calling~\cite{function_calling}) can be uniformly described as a backend-spec item, and the static-structure applications can also be covered by a PDGraph where the jumping probability between any two units is always 1.
In particular, for each LLM unit, we record the input/output length as well as the request parallelism instead of the absolute time to enable adaption of diverse runtime execution platforms (e.g., A100 and H100). 
Those properties can facilitate multiple efficiency optimization aspects: % queuing optimization and backend prewarming:
recording the backend consumptions can facilitate the prioritization of short applications;
recording the backend specifics and the jumping relationship can help prewarm the soon-called backends.

%\yfliu{
Moreover, to handle dynamic backend consumptions and probabilistic jumping relationships, we express the profiled resource demand as a distribution instead of as a single value. 
%We then elaborate on how to capture the dynamic demand of each functional unit with diverse backends. 
%Meanwhile, when expressing the probabilistic demand in each dimension (e.g., input/output token length and request parallelism), 
To be specific, after each profiling run, we append the execution information of each functional unit to a list. 
%This approach better supports our later usage compared to analytic expressions\footnote{
For dynamic backend consumptions, we note that recording the raw values is better than recording the coefficients of the fitted skewed norm distribution due to the depiction fidelity (the true distribution is in fact irregular) and computation efficiency: it takes much time (up to seconds) to fit out the coefficients as well as to support our later calculations. % (e.g., correlation analysis and integral calculus).
For probabilistic jumping relationships, we calculate the historical jumping frequency as the branch-taken probability.
Meanwhile, we set the maximal number of recorded values to $1000$ (evicted in FIFO manner), and the storage cost is indeed negligible.

Further, we use the Monte Carlo method to estimate the total demand of the entire application. 
To be specific, we perform random walk along the PDGraph (sampling the downstream branch as well as the unit demand value), until a sufficient number of samples are collected.
Note that this can be done efficiently because of the limited number of nodes and list sizes in typical PDGraphs.
by summing up the LLM execution time (input/output token lengths are transferred to the absolute service time based on the average per-token processing time in the runtime environment) and non-LLM execution time, we can get an estimation of the total execution cost for an application. 

%\yfliu{
% It needs to be noted that PDGraph itself is not designed to directly predict the determinate execution time, but rather to capture the variability of an LLM application in the form of distributions in various demand dimensions. Thus, directly evaluating its prediction accuracy is inapplicable. Alternatively, we indirectly evaluate the performance superiority brought by PDGraph in \Cref{subsec:ablation}, which confirms that our proposed correlation-aware refinement and Gittins-index calculation---leveraging the information recorded in PDGraph---can effectively reduce the average application completion time.
% }

\phm{Online estimation refinement for better accuracy.}
Accurate demand estimation is crucial for the overall serving efficiency of LLM applications. 
Yet, naively using the historical execution is insufficient: it is merely a prior knowledge and the runtime execution information---the posterior knowledge---is also valuable for refining the estimation results. 
In fact, due to the structural dependencies among stages, the resource demands of a stage are often correlated to the upstream ones. 
Specifically, we note that there typically exist three cross-unit demand correlation patterns:
\begin{itemize}
    \item  A unit's input length may correlate to the upstream unit's input/output length.
    For example, for the DM application, the input of each request in the \texttt{scoring} unit is a superset of a request's output in the (upstream) \texttt{aggregate} unit (plus a fixed system prompt). 
    Meanwhile, for the looping unit, downstream requests would share the same prompt template as the upstream ones, indicating similarity in the input length.
    \item A unit's output length may be correlated to its input length as well as the upstream unit's output length.
% Note that the input and output length might be correlated in some cases.
For example, for the \texttt{generate-code} unit in CG application, a more complex input task—typically associated with a longer prompt—tends to produce a longer code segment. 
% This pattern also applies to other LLM tasks, such as translation and article correction or polishing.
% Meanwhile, for the looping states, the request output lengths are also similar across units.
Similarly, the output lengths of requests are also similar across looping units.
    \item A unit's parallelism may be correlated to the upstream unit's parallelism. 
For example, as shown in KBQAV applications, for each inference request in the (upstream) \texttt{generate-queries} unit, a corresponding inference request would be launched at the (downstream) \texttt{verify-claim} unit.
\end{itemize}
% \emph{1)}
    
% \emph{2)} 

% \emph{3)} 

\begin{figure}[t]
\centering
\vspace{-0.1in}
    \subfloat[KBQAV]
    {	
        \includegraphics[width=0.4\linewidth]{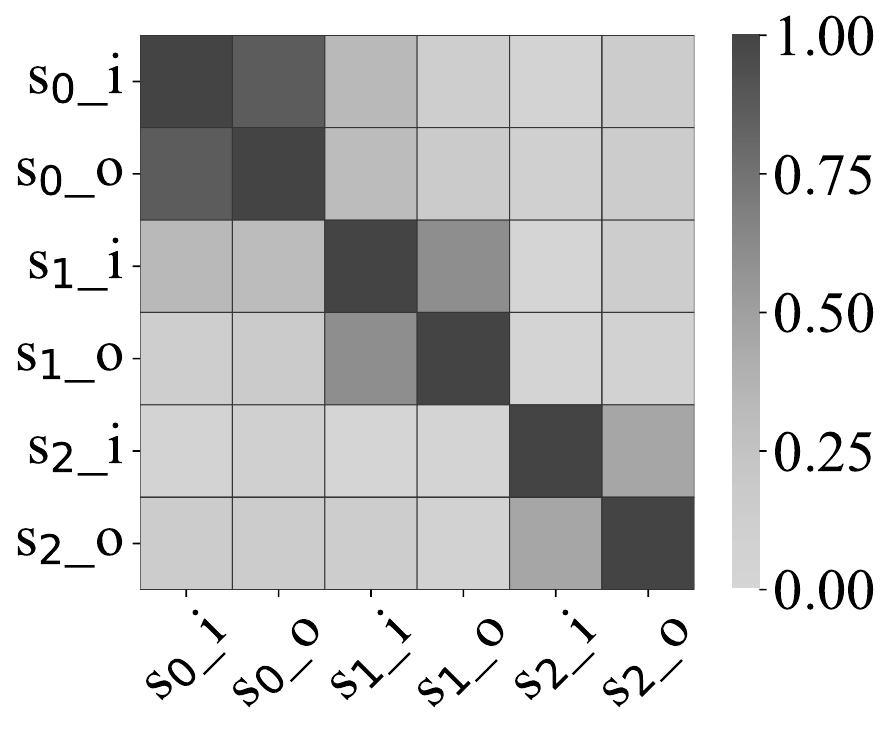}
        \label{fig:heatmap_kbqa}
    }\hfill  
    \subfloat[Document Merging (DM)]
    {
        \label{fig:heatmap_docmerge}
        \includegraphics[width=0.4\linewidth]{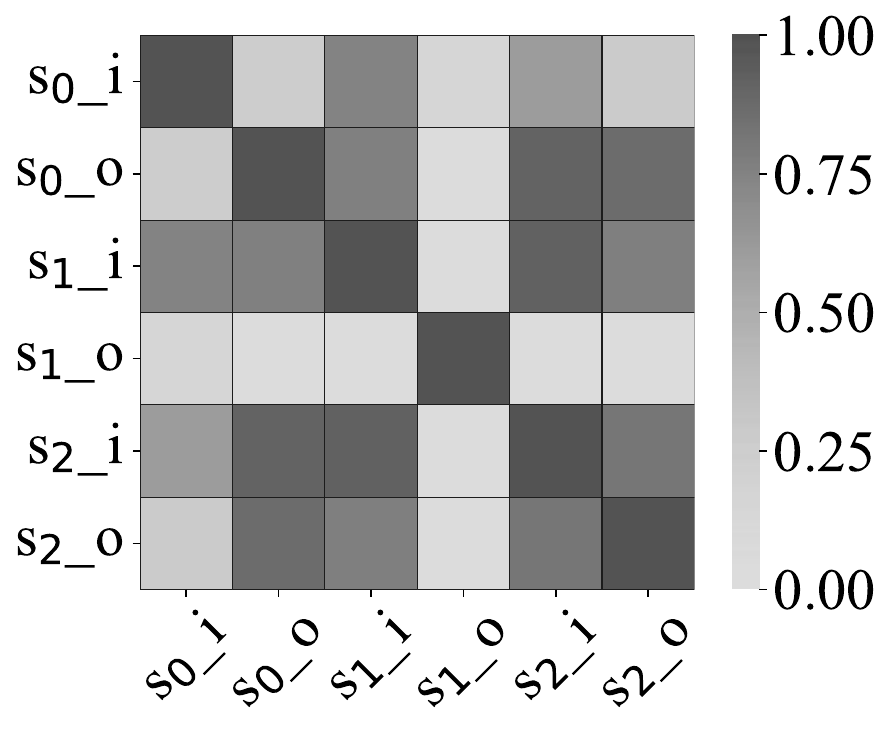}
    }\hfill    

	\vspace{-0.05in}
	\caption{The Pearson correlation coefficients between demands of dependent functional units. We show three units respectively in KBQAV and Document Merging, where $s_x$\_$i$, $s_x$\_$o$ represent the input/output token length of the $x$-th unit. 
        }
	\vspace{-0.1in}
	\label{fig:heatmap}
\end{figure}

To verify the existence of the above correlation types, we resort to \emph{Pearson Correlation Analysis}~\cite{sedgwick2012pearson} based on our profiled data over 100 trial runs.
We divide the demand range into 10 buckets and use $P(X=i) (i=0,1,…,9)$ to denote the probability that the demand quantity variable $X$ falls into bucket-$i$.
Then we calculate the Pearson correlation coefficient between two demand variables X and Y as
% Although modeling revealed that the input and output lengths of LLM requests approximate a skewed distribution, tending to cluster around the mean, the above experiments indicate that relying solely on the offline profile's mean is insufficient to achieve satisfactory prediction accuracy. Therefore, we aim to incorporate an online updating mechanism. Additionally, during the experiments, it was observed that the input and output lengths of different stages are directly or indirectly related. For instance, in React-style tasks, the output from one stage is directly added to the input of the next, leading to a growth trend in inputs. Similarly, in document merge tasks, longer inputs at the beginning often result in longer inputs and outputs in subsequent stages. We measured the correlation between stages of two applications using the Pearson correlation coefficient and represented it with the heatmap Figure \ref{fig:heatmap}. From the visualization results, it is evident that there is indeed a correlation between the inputs and outputs of certain stages within the app. Thus, we introduce Bayesian networks to learn such relationships. 
$\rho_{X,Y} = \frac{\text{cov}(X,Y)}{\sigma_X \sigma_Y} = \frac{\mathbb{E}((X - \mu_X)(Y - \mu_Y))}{\sigma_X \sigma_Y}.$	
In Fig.~\ref{fig:heatmap}, for KBQAV and Document Merging, we further depict the Pearson correlation coefficients between any two demand variables (each represents the input/output length of a unit).
For KBQAV, each unit's output length is strongly correlated to its input length; yet for DM, each unit's input length is highly-related to the upstream unit's input length. 

Fig.~\ref{fig:heatmap} also suggests that each application has distinct correlation patterns; for prediction efficiency, we only consider the demand correlations with a coefficient ($\rho$) larger than 0.5.
To that end, in each unit we add a five-tuple ($M_I^{\tilde{I}}$, $M_I^{\tilde{O}}$, $M_O^{\tilde{O}}$, $M_O^I$, $M_P^{\tilde{P}}$) to mask whether the corresponded demand correlation holds: $M_Y^X$ represents whether the demand variable $X$ affects $Y$; $I$, $O$ and $P$ (${\tilde{I}}$, ${\tilde{O}}$ and ${\tilde{P}}$) respectively represents the input length, output length and request parallelism of the current (upstream) unit.

The above correlation analysis enables more precise online demand prediction.
Upon the completion of a unit, its execution information can be immediately adopted for demand prediction of the future units. 
In that case, we are facing a \emph{conditional prediction} problem.
Suppose the request input and output length of the just-finished unit is respectively in bucket $i$ and $o$---with both correlated with the input length of the downstream unit, and we need to predict $P(I|\tilde{I}=I, \tilde{O}=o)$. 
To accomplish that, we join the historical execution records of the two dependent functional units (i.e., yielding $(\tilde{I}, \tilde{O}, I)$ tuples respectively from each trial round), and filter out the profiled tuples satisfying $\tilde{I}=i$ and $\tilde{O}=o$; then the $I$ distribution within the filtered records, which guides our Monto Carlo sampling process, can yield a more accurate estimation of the resource demand in the current round.

\subsection{PDGraph-based Queue Management}
\label{subsec:scheduling}

With PDGraph, we can now obtain a probabilistic demand estimation of an LLM application's overall resource demand. 
In this part, we explore how to leverage such demand estimation to optimize the overall queuing performance.

\phm{Optimizing average ACT based on the \emph{Gittins} policy.}
Regarding efficient scheduling of LLM applications, a primary metric is the average \emph{application completion time} (ACT). 
%we first focus on optimizing the average application completion time. 
A classical algorithm proven optimal in that regard is \emph{shortest-remaining-processing-time} (SRPT)~\cite{schrage1968proof}.
%, which always prioritizes the application with the rho each request based on the remaining processing time.
However, naively applying SRPT is not appropriate for our problem: SRPT relies on deterministic demand knowledge to make scheduling decisions, yet the resource demands of LLM applications are uncertain a priori (our PDGraph strategy mitigates but does not totally eliminate such uncertainty). 
Although it is possible to use the expected value (mean) of a demand distribution to emulate SRPT scheduling, this method often fails to work well due to its blindness to instantaneous execution progress.		
For example, for a request with an expected output length of 20 tokens, it is possible that its token generation process does not stop even after 100 tokens.
In that case, we need to timely refresh the demand estimation based on the latest progress, rather than sticking to the prior expectations. Otherwise, the remaining processing time---the expected execution time minus the executed time---may ironically become \emph{negative}. 
Therefore, we need to introduce uncertainty-awareness in designing our scheduling algorithm.

\begin{figure}
    \centering
    \includegraphics[width=1\linewidth]{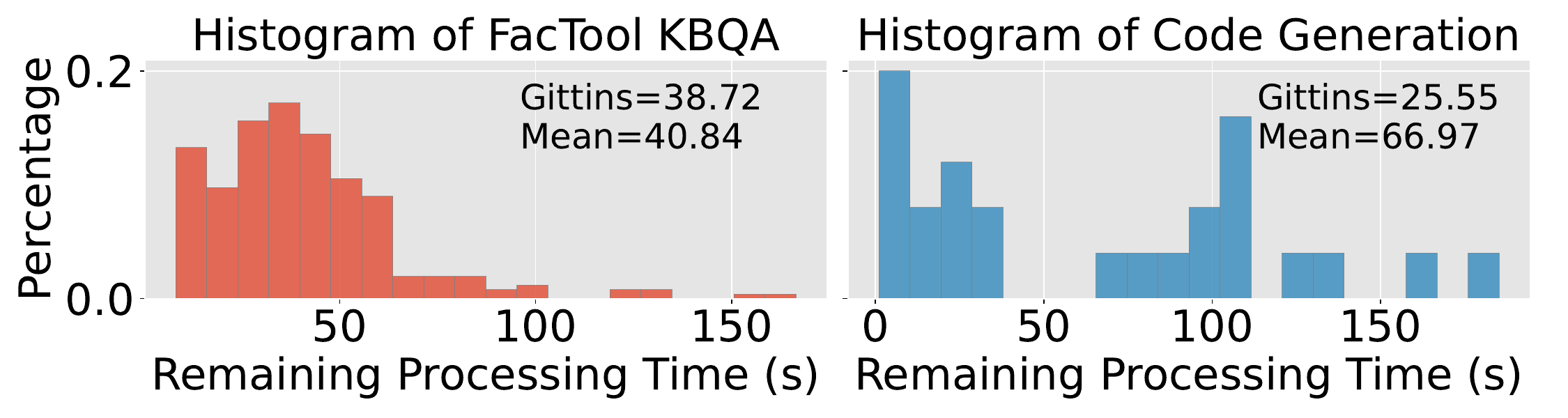}
    \caption{
    % Given two LLM applications, Math\_FacTool and ALFW\_ReAct, under SPRT it is the former that is scheduled first, yet under the Gittins policy it is the latter (with a lower Gittins rank).
    Given two running LLM applications, FacTool\_KBQA and Code\_Generation, under SPRT it is the former that is scheduled first, yet under the Gittins policy it is the latter (with a lower Gittins rank).
    % Regarding the probability to complete within three additional iterations, for Math\_FacTool it is 7\%, yet for ALFW\_ReAct it is 48\%; this suggests that prioritizing ALFW\_ReAct indeed has a higher reward in practice. 
    Regarding the probability to complete within 9 seconds, for FacTool\_KBQA it is 0.3\%, yet for Code\_Generation it is 16\%; this suggests that prioritizing Code\_Generation indeed has a higher reward in practice. }
    %\chen{request or app?}
    %Illustration of the shortcomings of using a simple distribution mean, which fails to provide effective insights for selecting tasks with higher reward rates.
    \label{fig:solution_gittins_histogram}
\end{figure}

The problem we now face---minimizing the average completion time for jobs with \emph{unknown durations} but \emph{known duration distributions}---has been studied in the literature, for which the \emph{Gittins} policy~\cite{gittins2011multi, aalto2009gittins, scully2021gittins} has been proven optimal.
The basic idea of Gittins policy is to calculate a \emph{Gittins rank} for each job---based on its executed time so far and its size distribution---as the scheduling priority.
% \yfliu{
% Specifically, we have built an application's duration distribution $\mathcal{D}$ using PDGraph, and it has been executed for a time period of $a$,
% }
Specifically, let an application's duration distribution be $\mathcal{D}$, and it has been executed for a time period of $a$,
% First, we calculate the Gittins index $g_{output}$ for the output length of each request within the coinference. 
% Let the output length distribution of a request be $d$, and suppose $a$ tokens have already been generated. 
then its Gittins rank $G$ can be expressed as: 
\begin{equation}
\label{eq:gittins_output}
    \begin{aligned}
    G(\mathcal{D},a) = \operatorname*{inf}_{\Delta>0}{\frac{\operatorname{E}[\operatorname*{min}\{X_{\mathcal{D}}-a,\Delta\}\mid X_{\mathcal{D}}>a]}{\operatorname*{P}\{X_{\mathcal{D}}-a\leq\Delta\mid X_{\mathcal{D}}>a\}}},
    \end{aligned}
\end{equation}
where $X_\mathcal{D}$ is the random variable under $\mathcal{D}$.
% G can be viewed as the estimated output given the latest execution situation:
Given a service budget $\Delta$ (i.e., the number of additional tokens allowed to be generated henceforth), the denominator in Eq.~\ref{eq:gittins_output} represents the possibility that the generation process can finish before $\Delta$ (which contributes to a small average completion time), and the numerator represents the corresponding resource consumption (the cost paid to serve it). 
To determine the scheduling priority, the Gittins Index essentially computes the maximum achievable cost-to-return ratio under any possible cost budget,
%given the current uncertainty (belief state $X_{\mathcal{D}}>a$).}
% Note that $G$ is always no larger than the maximum value in $\mathcal{D}$, and it can be viewed as the \emph{true runtime estimator of the remaining processing demand} when determining the scheduling priority:
and existing works~\cite{gittins2011multi, scully2021gittins} have shown that this method can theoretically yield the minimal average completion time.
In Fig.~\ref{fig:solution_gittins_histogram}, we show a case where the Gittins policy yields a different yet better scheduling order than SPRT.
Besides, for computation efficiency, in practice we update the estimation of Gittins index only after each bucket period. % the bucket granularity. 

\phm{Extend to other efficiency\footnote{Note that, while we primarily focus on the efficiency aspect of LLM application scheduling in this paper, with the demand information from PDGraph, we can in fact also enhance the fairness aspect by enabling demand-aware fair scheduling methods like weighted fair queuing~\cite{bennett1996wf}.} objectives.}
Regarding efficient LLM application scheduling, apart from the ACT criterion, 
%some other criteria are also common in the literature. 
% In some cases, 
% \chen{describe in greater detail}
users may also associate a deadline with their application request, and expect that the final output can be returned before that deadline.
In that cases, the efficiency criterion is the ratio of applications that can satisfy the deadline requirements.
%With our PDGraph modeling, we note that
For deadline-based scheduling, we note that PDGraph-based demand modeling can also help to attain better performance---surpassing the demand-agnostic scheduling methods exemplified by Earliest-deadline-first (EDF)~\cite{andrews2000probabilistic}.
%On the one hand, 
EDF always prioritizes the application with the earliest deadline, yet an earlier deadline does not mean a higher urgency---it also depends on the demand volume. 
% Based on probabilistic demand modeling, we can calculate the risk that an LLM application---if scheduled instantly---finishes after its specified deadline.
{
Based on PDGraph, we define the worst-case slack time $S$ relative to the deadline as
\begin{equation}
\label{eq:ddl_violation_risk}
    \begin{aligned}
    S(\mathcal{D},a) = t_{\text{ddl}} - t_{\text{now}} - \left( \sup X_{\mathcal{D}} - a \right),
    % r_\text{ddl} = ddl - now - (\sup X_{\mathcal{D}} - a)
    % r_\text{ddl} = \frac{\sup X_{\mathcal{D}} - a}{ddl - now},
    \end{aligned}
\end{equation}
% \begin{equation}
% \label{eq:ddl_violation_risk}
%     \begin{aligned}
%     R(\mathcal{D},a) = 
%     \frac
%     {\operatorname*{P}\{X_{\mathcal{D}}-a > ddl - now \mid X_{\mathcal{D}}>a\}}
%     {\operatorname{E}[ X_{\mathcal{D}}-a > ddl - now \mid X_{\mathcal{D}}>a]},
%     \end{aligned}
% \end{equation}
% where the numerator quantifies the probability of historical samples missing the deadline (with higher values indicating elevated risk), while the denominator captures the time remaining until a potential violation (where smaller values denote increased urgency).
where $S$ represents the worst-case remaining time till deadline.
We then prioritize applications in ascending order of $S$, emulating the \emph{Least Slack Time First} (LSTF) policy~\cite{davis1993scheduling}.
}

\subsection{PDGraph-based Backend Prewarming}
\label{subsec:provisioning}

\begin{figure}
    \centering
    \vspace{-0.1in}
    \subfloat[Effective Prewarming]{
        \includegraphics[width=0.532\linewidth]{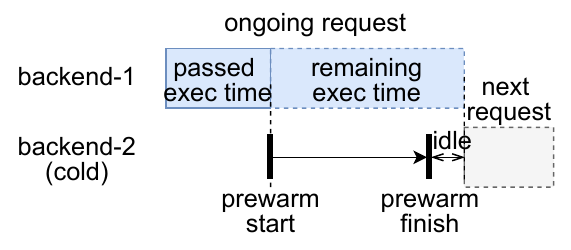}
    }
    \subfloat[Ineffective Prewarming]{
        \includegraphics[width=0.42\linewidth]{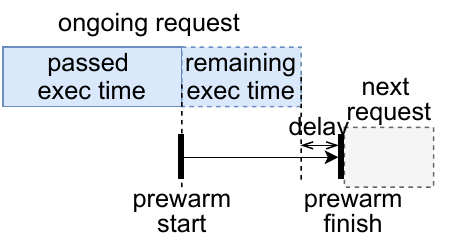}
    }
    
    \caption{Trade-off in setting up the prewarming triggering time: prewarming too early wastes resources, yet prewarming too late delays application completion.}
    \label{fig:example_prewarm}
    \vspace{-0.1in}
\end{figure}

As elaborated in \Cref{subsec:limitation}, cold backends would delay the completion of an LLM application. 
With PDGraph, Hermes can estimate the arrival of the downstream requests as well as their desired backend types. 
By prewarming cold backend to be used by the downstream unit before the completion of the current unit, we can get rid of the warm-up delay from the critical path. 
However, backend prewarming is not free-of-charge. 
On the one hand, a functional unit may have multiple downstream units, and it is possible that the prewarmed backends are never used; on the other hand, the current unit may complete much later than expected, meaning that the prewarmed backend would wait idly for a long time.
Therefore, as shown in Fig.~\ref{fig:example_prewarm}, there exists a trade-off in determining the prewarming triggering moment:
more aggressive prewarming can help attain faster application completion---yet at the cost of larger resource wastage. 

We then introduce a knob $K$, called \emph{expected prewarming effectiveness}, to tune that trade-off. 
To be specific,  given a running function unit, if one of its downstream backends is not active, we then determine \emph{whether} and \emph{when} to trigger prewarming with an analytical method.
We let $p_s$ be the probability of selecting the target unit, $t_c$ be the actual completion time of the current unit, $t_s$ be the moment to start prewarming, and $t_p$ be the duration of the prewarming operation.
Then the probability that prewarming is effective (i.e., when the target downstream request finally arrives, its backend is already well-warmed) can be expressed by 
\begin{equation}
p_e=p_s*P(t_c>t_s+t_p).
\end{equation}
Given the knob $K$, if $p_s<K$, then Hermes does not trigger prewarming; otherwise, it triggers prewarming at the time such that  $p_e=K$.
Such a knob $K$ can be deemed as a kind of service level agreement: a more significant application (e.g., from premium users) can be assigned with a smaller $K$ value. 

The above prewarming technique applies to the non-LLM backends like docker and DNNs, and also applies to the cache resources in LLM backends.
In fact, the cache space on LLM servers is significant for efficient LLM serving. 
% Caching is significant for efficient LLM serving. 
Existing works~\cite{kwon2023efficient,abhyankarinfercept} have shown that KV-cache can remarkably improve the token generation speed for requests sharing the same prompt tokens.
Meanwhile, Low-Rank Adaptation (LoRA)~\cite{hu2021lora} is prevalently adopted to support LLM serving with customized models;
%, which, in production clusters with diverse LoRA adaptors, requires runtime adaptor loading to multiplex the base model~\cite{sheng2023s}.
those LoRA adaptors take non-negligible time to load and should thus be also cached in memory~\cite{li2024caraserve} (\ref{subsec:limitation}). 
In that sense, preloading the desired KV cache or LoRA is also necessary for the efficient execution of LLM applications, which will also be evaluated later in \Cref{subsec:eval_backend_quality}.%to the estimated arrival time of downstream unitsthe cache space may also be contended by multiple applications.

\section{Implementation}
\label{sec:implementation}
We have implemented Hermes with 4100 lines of Python codes. 
% The profiled application execution information is stored in a JSON file called \texttt{KnowledgeBase}, and as shown in Fig.~\ref{fig:overview}, the key components we implemented are the \texttt{AppHandler} and the \texttt{HermesScheduler}.
As illustrated in Fig.~\ref{fig:overview}, our implementation comprises two key components: the \texttt{HermesScheduler} and the \texttt{HermesLet}. 
We uses vLLM 0.4.3~\cite{kwon2023efficient} as the low-level inference engine and ZeroMQ-based RPC~\cite{ZeroMQ} to exchange control messages between the \texttt{HermesScheduler} and the \texttt{HermesLet}.
Additionally, we build a comprehensive benchmark suite consisting of 11,300 lines of Python code, which includes all of the application types depicted in Fig.~\ref{fig:app_structure}. 

In building the PDGraph models, we profile each application for 1000 times, and store their PDGraphs in a JSON file. 
The average JSON file size for one application is around 100KB.
After loading such JSON file, the \texttt{HermesScheduler} launches a background RPC thread to keep refreshing the demand estimation (at per-bucket granularity) based on the latest execution status reported from the \texttt{HermesLet}.  
The \texttt{HermesScheduler} would accordingly refresh the application scheduling priority (based on the Gittins index or worst-case slack time) and notify the \texttt{HermesLet} once the priority changes. 
The \texttt{HermesScheduler} also issues backend prewarming signal at desired time to the \texttt{HermesLet}.
The \texttt{HermesLet} also maintains a dedicated background RPC client for continuous communication with the \texttt{HermesScheduler}, through which it reports execution status while receiving updated request priorities and prewarming signals.
It also employs two dedicated background threads to prefetch KV cache blocks 
% with layer-wise loading 
or LoRA adapters as well as to prewarm docker containers, in accordance with the \texttt{HermesScheduler}'s decisions.
\section{Evaluation}
\label{sec:eval}

\subsection{General Setups}

\begin{figure*}[t]
    \centering
    \begin{minipage}[t]{0.48\linewidth}
        \subfloat[The average ACT with varying application arrival intensities. ]{
            \label{fig:evaluation_act_a100}
            \centering
            \includegraphics[width=0.9\linewidth]{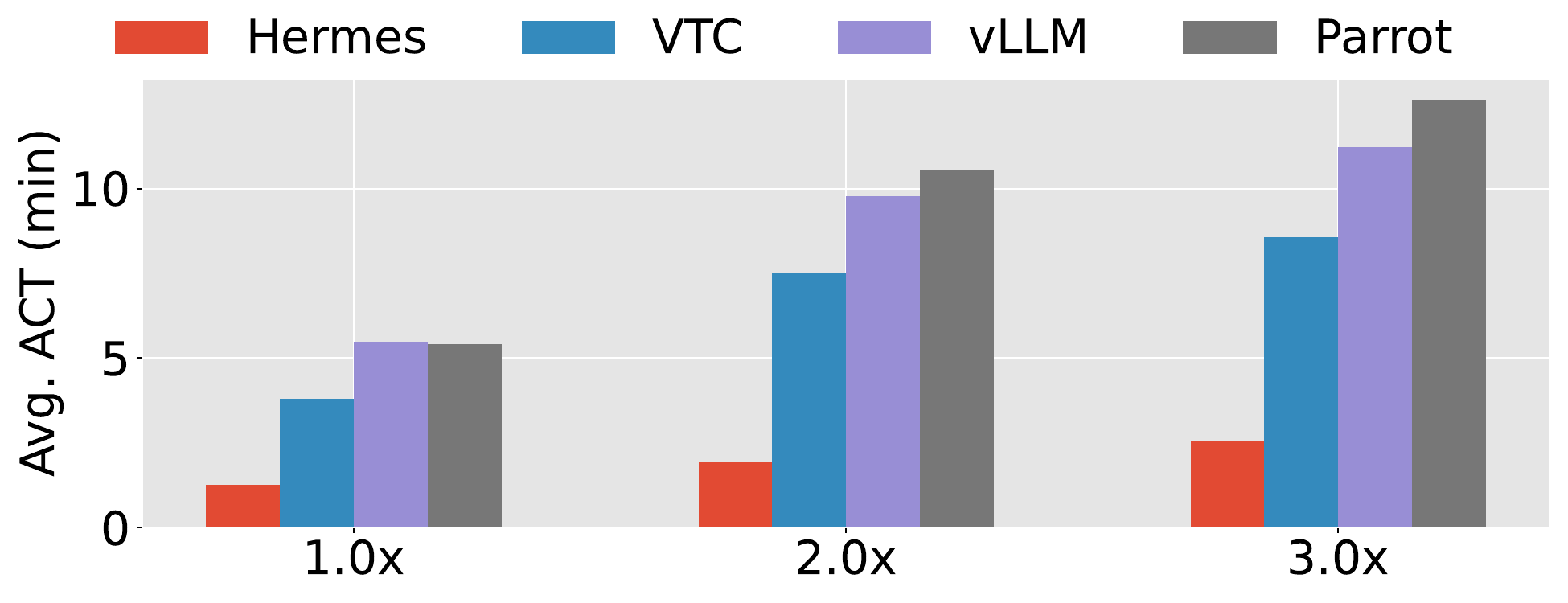}
        }
        \hfill
        \subfloat[The CDF of ACTs under different schedulers. ]{
            \label{fig:evaluation_cdf_a100}
            \centering
            \includegraphics[width=0.9\linewidth]{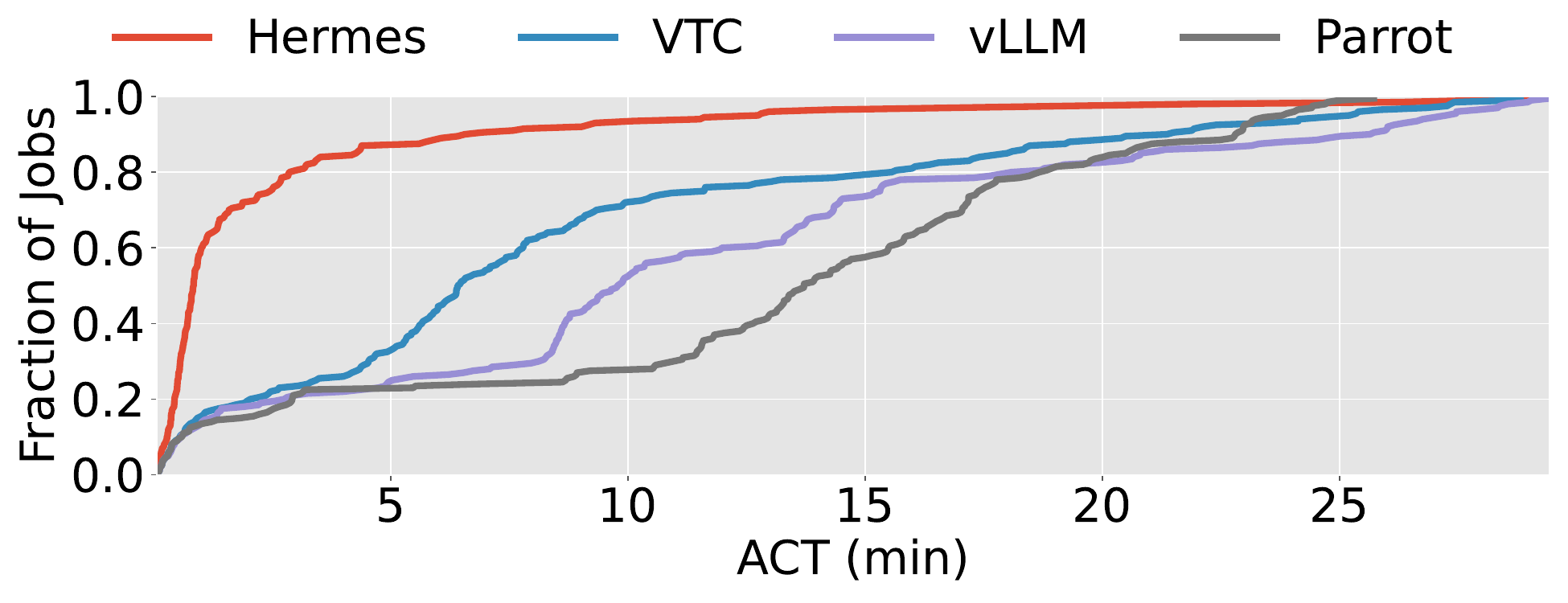}
        }
        \caption{The experimental results from a single LLaMA2-7B model running on an A100 GPU. %The overhead of the Hermes scheduler during each scheduling round and the updates triggered upon application completion.
        }
        \label{fig:evaluation_a100}
    \end{minipage}
    \hfill
    \begin{minipage}[t]{0.48\linewidth}
        \subfloat[The average ACT with varying application arrival intensities. ]{
            \label{fig:evaluation_act_h800}
            \centering
            \includegraphics[width=0.9\linewidth]{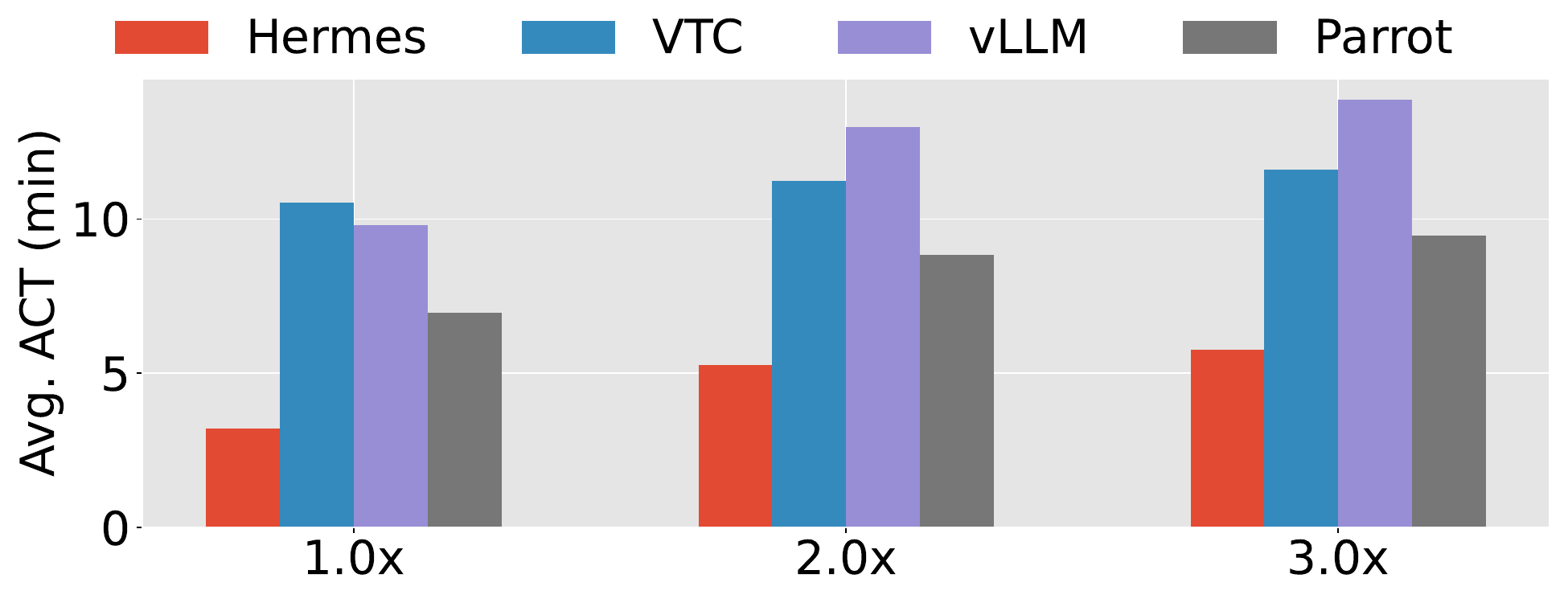}
        }
        \hfill
        \subfloat[The CDF of ACTs under different schedulers. ]{
            \label{fig:evaluation_cdf_h800}
            \centering
            \includegraphics[width=0.9\linewidth]{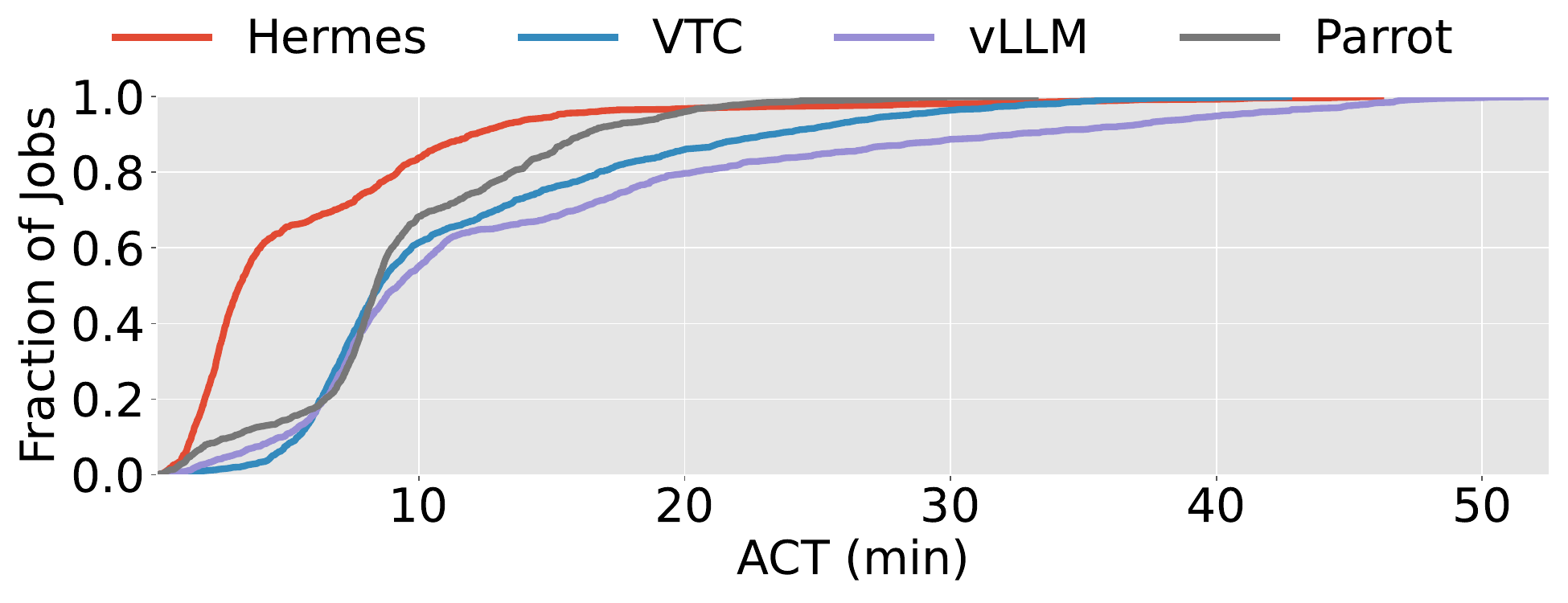}
        }
        \caption{The experimental results from two LLaMA3-70B models running on eight H800 GPUs. %The overhead of the Hermes scheduler during each scheduling round and the updates triggered upon application completion.
        }
        \label{fig:evaluation_h800}
    \end{minipage}
    % \hfill
    % \begin{minipage}[t]{0.32\linewidth}
    %     \centering
    %     \includegraphics[width=1\linewidth]{figures/evaluation_ddl.pdf}
    %     \caption{The DDL Satisfactory Ratio under different schedulers.
    %     }
    %     \label{fig:evaluation_cdf}
    % \end{minipage}
    % \hfill
    % \begin{minipage}[t]{0.32\linewidth}
    %     \centering
    %     \includegraphics[width=1\linewidth]{figures/evaluation_sched.pdf}
    %     \caption{
    %     Ablation study on the methods tackling demand dynamicity.
    %     }
    %     \label{fig:evaluation_sched}
    % \end{minipage}
\end{figure*}

% \begin{figure*}[t]
%     \centering
%     \begin{minipage}[t]{0.32\linewidth}
%         \centering
%         \includegraphics[width=1\linewidth]{figures/evaluation_act.pdf}
%         \caption{The average ACT with varying application arrival intensities.
%         }
%         \label{fig:evaluation_act}
%     \end{minipage}
%     \hfill
%     \begin{minipage}[t]{0.32\linewidth}
%         \centering
%         \includegraphics[width=1\linewidth]{figures/evaluation_cdf.pdf}
%         \caption{The CDF of ACTs under different schedulers.
%         }
%         \label{fig:evaluation_ddl}
%     \end{minipage}
%     \hfill
%     \begin{minipage}[t]{0.32\linewidth}
%         \centering
%         \includegraphics[width=1\linewidth]{figures/evaluation_ddl.pdf}
%         \caption{The DDL Satisfactory Ratio under different schedulers.
%         }
%         \label{fig:evaluation_cdf}
%     \end{minipage}
%     \hfill
%     \begin{minipage}[t]{0.32\linewidth}
%         \centering
%         \includegraphics[width=1\linewidth]{figures/evaluation_sched.pdf}
%         \caption{
%         Ablation study on the methods tackling demand dynamicity.
%         }
%         \label{fig:evaluation_sched}
%     \end{minipage}
% \end{figure*}

\begin{figure}[t]
    \centering
    \begin{minipage}[t]{0.49\linewidth}
        \centering
        \includegraphics[width=1\linewidth]{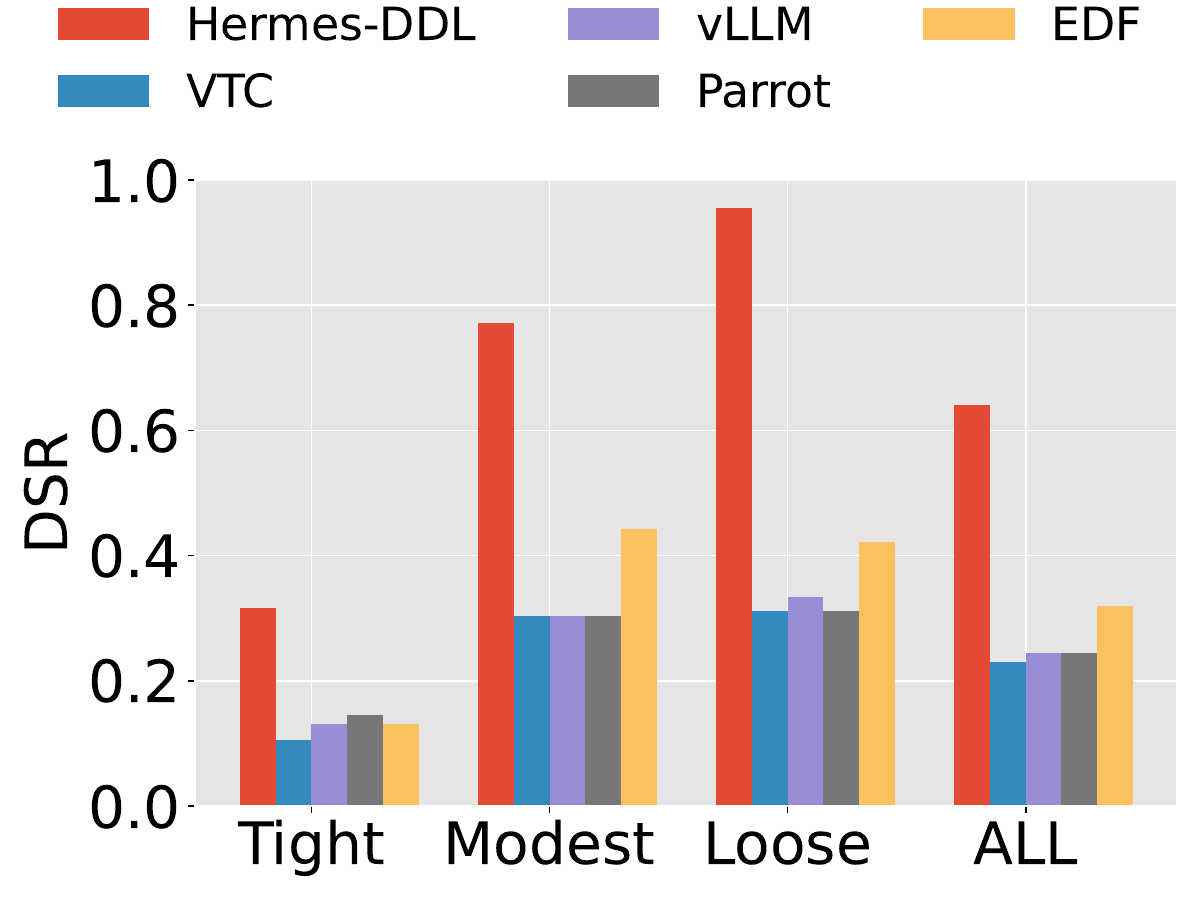}
        \caption{The DDL Satisfactory Ratio under different schedulers.
        }
        \label{fig:evaluation_ddl}
    \end{minipage}
    \hfill
    \begin{minipage}[t]{0.49\linewidth}
        \centering
        \includegraphics[width=1\linewidth]{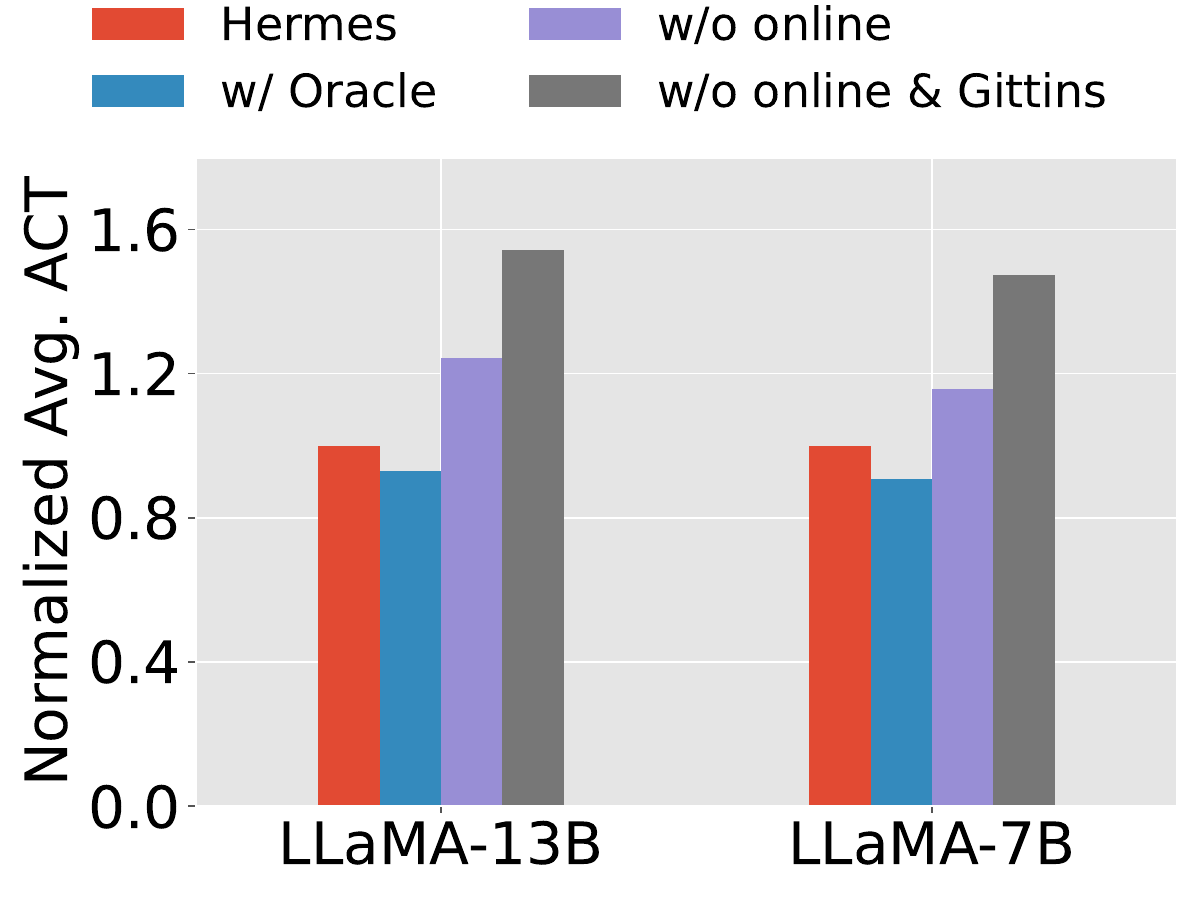}
        \caption{
        Ablation study on the methods tackling demand dynamicity.
        }
        \label{fig:evaluation_sched}
    \end{minipage}
\end{figure}

% \begin{figure}
%     \centering
%     \includegraphics[width=\linewidth]{figures/evaluation_ddl.pdf}
%     \caption{The DDL Satisfactory Ratio under different schedulers.}
%     \label{fig:evaluation_ddl}
% \end{figure}

% \begin{figure}
%     \centering
%     \includegraphics[width=\linewidth]{figures/evaluation_sched.pdf}
%     \caption{Ablation study on the methods tackling demand dynamicity.}
%     \label{fig:evaluation_sched}
% \end{figure}

\phm{Hardware platform.}
We evaluate Hermes respectively with single- and multi-GPU experiments.
The single-GPU evaluations use a server with one A100 GPU, and the multi-GPU evaluations use a server with eight H800 GPUs. 
The single-GPU server has four 16-core AMD EPYC 7302 CPUs, 128GB DRAM, 4TB SSDs and one NVIDIA A100-PCIe-40GB GPU. 
The multi-GPU server has four 48-core Intel Xeon Platinum 8558 CPUs, 2TB DRAM, 28TB SSDs and eight NVIDIA H800-80GB GPUs. 
% Both servers run CUDA 12.4.
Unless otherwise specified, we deployed LLaMA2-7B and LLaMA2-13B on a single A100-PCIe-40GB GPU, and two LLaMA3-70B models across eight H800-80GB GPUs using 4-way tensor parallelism to further evaluate Hermes' performance at scale.

\phm{Workloads.}
To investigate the performance of Hermes in real-world environments, we sampled the request arrival time distribution from the trace published by MoonCake~\cite{qin2024mooncake}, from which we can clearly observe the bursty arrival patterns in real-world scenarios. 
%Since the MoonCake trace does not contain LLM application-specific information, 
We utilized its arrival time distribution while sampling real inputs from the application families shown in Fig.~\ref{fig:app_structure}.
% For our experiments, we create a mixed workload set with totally 300 LLM applications (each with distinct inputs from the original datasets). % from diverse categories.
% In that workload suite, 20\% applications are multi-turn conversations requiring guaranteed TPT performance (TPT-sensitive), 20\% require meeting a preset deadline (DDL-sensitive), and the remaining 60\% target for short ACT (ACT-sensitive). 
% To get the applications not involving TPT objectives, we randomly sample from the 9 non-conversation application families in Fig.~\ref{fig:app_structure}.
% In particular, similar to prior work~\cite{qiao2021pollux, jayaram2023sia, zheng2023shockwave}, we set the sampling probability of small (EV, FEV, CC, ALFWI and KBQAV---usually less than 1 min), medium (CG and PE---usually between 1 and 10 min) and large (DM and MRS---usually longer than 10 min) applications to be 72\%, 26\%, and 2\%, respectively.
Specifically, similar to prior work~\cite{qiao2021pollux, jayaram2023sia, zheng2023shockwave}, we set the sampling probability of \emph{small} (EV, FEV, CC, ALFWI and KBQAV---usually less than 1 min), \emph{medium} (CG and PE---usually between 1 and 10 min) and \emph{large} (DM and MRS---usually longer than 10 min) applications to be 72\%, 26\%, and 2\%, respectively.
% Meanwhile, for the TPT-sensitive applications, we set the TPT target as 0.1 s/token (given that typical human reading speed is less than 600 token per minute~\cite{reading_speed}).
% For the DDL-sensitive applications, we set the deadlines by multiplying the original execution length by a random factor between 1.2 and 2, resembling some previous studies~\cite{narayanan2020heterogeneity, gu2023elasticflow}.
% Regarding application submission, given a submission window of T, we let the submission interval follow a Poison distribution with the expectation be T/N (N is the total application number submitted in experiments). 

\phm{Baselines.}
We compare Hermes with three scheduling strategies: vLLM~\cite{vllm}, Parrot~\cite{lin2024parrot}, and VTC~\cite{sheng2024fairness}.
As explained in \Cref{subsec:limitation}, vLLM adopts the FCFS policy at the request level, and Parrot adopts the FCFS policy at the application level.
VTC seeks to fairly serve the tasks from different users. %, which is also agnostic to the application demands.
%vLLM improves memory utilization through PagedAttention, but it schedules requests in a First-Come-First-Serve (FCFS) manner, lacking awareness of upper-layer LLM application requirements. 
%Parrot performs FCFS scheduling at the application level, meaning that each request is scheduled based on the arrival time of the coinference it belongs to.
%VTC prioritizes requests from applications that have received the least service, thereby ensuring fairness among applications. Compared to the FCFS algorithm, it achieves relatively higher efficiency.
For scenarios with deadlines, we additionally include the Earliest-Deadline-First (EDF) policy~\cite{andrews2000probabilistic} as a baseline.
We defer the introduction of other baselines to each specific micro-benchmark experiment.
Regarding the default hyperparameter setup in Hermes, the threshold on Pearson correlation coefficient (\Cref{subsec:demand_modeling}) is set to 0.5, 
%\chen{the threshold on deadline-violation risk (\Cref{subsec:scheduling}) is set to 0.25---do we use this? if yes a sensitivity analysis is necessary; otherwise we simply prioritize the app with the highest risk?}\yfliu{Current evaluations either consider only act app or only ddl app.}, 
the expected prewarming effectiveness knob $K$ (\Cref{subsec:provisioning}) is set to 0.5, and the number of buckets for distribution description (\Cref{subsec:demand_modeling} and \Cref{subsec:provisioning}) is set to 10. 

% \noindent\textbf{Workloads.}
% We developed a benchmark suite that includes the 10 workloads shown in Fig.~\ref{fig:app_structure}. 
% In our constructed end-to-end mixed workload, 20\% of the total submitted tasks in the trace are TPT-sensitive applications (specifically multi\_conversation), 20\% are DDL-sensitive applications, and 60\% are ACT-sensitive applications.
% Among non-TPT-sensitive applications, we categorized applications based on their execution time, and similar to prior work~\cite{qiao2021pollux, jayaram2023sia, zheng2023shockwave}, we set the probability of generating Small (0-1 min), Medium (1-10 min), and Large ($>$10 min) applications to be 72\%, 26\%, and 2\%, respectively.
% Small applications include Factool\_Generate-Code-Solution, Factool\_Knowledge-Based-QA, Factool\_Check-Math-Solution, ReAct\_Fact-Extraction-and-Verification, and ReAct\_ALFWorld; Medium applications include AutoGen\_CodeFeedback and HuggingGPT; and Large applications include LangChain\_MapReduce and GoT\_MergeDocuments. 
% The DDL values are determined by multiplying the original execution length by a random factor between 1.2 and 2, as done in previous studies~\cite{narayanan2020heterogeneity, gu2023elasticflow}.
% Following previous work~\cite{sheng2023s}, We generated application arrival times using a power-law distribution, with varying arrival rates represented by the exponent $\alpha$.

\subsection{End-to-end Scheduling Performance}
\label{subsec:end-to-end}

\phm{Minimizing average ACT.}
We first evaluate how Hermes can improve the application serving efficiency measured by the average application completion time.
We conducted two sets of experiments, running LLaMA2-7B and LLaMA3-70B respectively, and submitted 300 and 3000 applications correspondingly based on different hardware computing capabilities.
We evaluated the performance across submission windows of 10, 15, and 30 minutes (corresponding to workload intensity levels of $3\times$, $2\times$, and $1\times$), and reported the ACT distribution.
% We submit all the 300 applications with the submission window respectively set to 10, 12, 15, 20, and 30 minutes (i.e., the workload intensity respectively be $3\times$, $2.5\times$, $2\times$, $1.5\times$ and $1\times$). 
As shown in Fig.~\ref{fig:evaluation_act_a100} and Fig.~\ref{fig:evaluation_act_h800}, Hermes can perform much better than vLLM, Parrot, and VTC in each case.
% For example, when workload intensity is 1$\times$, the ACT performance of Hermes is 77.0\% (76.7\% and 66.7\%) better than vLLM (Parrot and VTC
% % \footnote{We note that VTC performs better than Parrot---because its fair-sharing policy can prevent short applications from being blocked by large ones.}
% ); when the workload intensity is $3\times$, the ACT performance of Hermes is 77.3\% (79.9\% and 70.3\%) better. % than vLLM (Parrot and VTC).
For example, in the LLaMA2-7B experiment under 1$\times$ workload intensity, the ACT performance of Hermes is 77.0\% (76.7\% and 66.7\%) better than vLLM (Parrot and VTC); in the LLaMA3-70B experiment under 3$\times$ workload intensity, the ACT performance of Hermes is 58.5\% (39.1\% and 50.3\%) better.
For tail application performance, in the LLaMA2-7B experiment under 1$\times$ workload intensity, the P95 ACT performance of Hermes is 82.4\% (69.0\% and 74.1\%) better than vLLM (Parrot and VTC); in the LLaMA3-70B experiment under 3$\times$ workload intensity, the P95 ACT performance of Hermes is 62.2\% (21.7\% and 45.8\%) better.

% In the end-to-end experiments, we run an LLM inference engine (LLaMA 7B) on a single A100 GPU as the backend server, while a frontend benchmark suite simulates application arrivals. We use a mixed workload consisting of 500 application submissions and vary the submission window (with a baseline of 30 minutes, reduced by 1-3x) to evaluate the average application completion times, deadline satisfaction ratios, and time-per-token satisfaction ratios under different workload intensities. Fig.~\ref{fig:evaluation_e2e} presents the overall performance between Hermes and baseline strategies. 

\phm{Maximizing DDL satisfactory ratio.} 
We next evaluate Hermes-DDL, a variant of Hermes specifically designed for deadline-constrained scenarios (based on methods in \Cref{subsec:scheduling}). 
Following the {LLaMA2-7B} experimental setup, we submit the same 300 applications within a 15-minute submission window. 
Additionally, we assign distinct deadlines to each task by scaling the original execution time with random factors of 1.2$\times$ (\emph{tight}), 1.5$\times$ (\emph{modest}), and 2$\times$ (\emph{loose}), consistent with methodologies employed in prior studies~\cite{narayanan2020heterogeneity, gu2023elasticflow}. 
%For this setup, we also include EDF as a baseline. 
Fig.~\ref{fig:evaluation_ddl} reports the Deadline Satisfaction Ratio (DSR, meaning the ratio of applications that can complete before the specified deadline)---for all the applications as well as for applications in each DDL-scaling category.
It shows that Hermes-DDL achieves the highest DSR among all the schemes evaluated, delivering a 1$\times$ improvement over EDF. 
This improvement primarily stems from Hermes-DDL's ability to leverage application demand information, with which it prioritizes the most urgent applications while deferring less critical ones. 

\subsection{Effect on Tackling Demand Dynamicity}
\label{subsec:ablation}

\phm{Setup.}
Recall that to address demand dynamicity, we have adopted multiple techniques including offline profiling, online refinement as well as the Gittins policy. 
We now verify their respective effectiveness by comparing the performance between four schemes: Hermes, Hermes without online refinement, Hermes without online refinement or Gittins, and Hermes with Oracle (representing the ideal case where the exact demands are available---by conducting a trial run with \texttt{temperature} set to 0).
% We first disable the correlation analysis module and get \emph{Hermes-v1}.
% Then we disable the Gittins policy (i.e., using the mean value of a distribution in SPRT scheduling), and further get \emph{Hermes-v2}. 
% For reference, we also add an idealized baseline, Hermes-Oracle, which knows exactly the resource demand of each LLM application (by conducting a trial run with \texttt{temperature} set to 0).
%In our experiments, we submit 300 applications to the A100 server within a time window of 10 minutes, and measure the average ACT (normalized by that under standard Hermes).
We conduct this experiment respectively with LLaMA-7B and LLaMA-13B as the LLM backend,
%LLaMA-7B uses two V100 GPUs with tensor parallel deployment, and LLaMA-13B uses four. 
and submit 300 applications within a time window of 10 minutes.

\phm{Results.}
Fig.~\ref{fig:evaluation_sched} shows the average ACT in each case (normalized by that under Hermes).
Without online correlation analysis, the average ACT would be inflated by 15.3\% with the LLaMA-7B model. 
When the Gittins policy is further disabled, that performance degradation would increase to 47.5\%.
Such results confirm the indispensability of both online refinement and the Gittins policy in tackling demand uncertainty.
Meanwhile, the performance gap between Hermes and Hermes-Oracle is less than 10\%, suggesting that Hermes can deliver near-optimal scheduling performance for LLM applications. %albeit confronting the dynamic demands of LLM applications.
We also notice that in each case, the performance of LLaMA-7B and LLaMA-13B are quite similar.

\subsection{Effect of Backend Prewarming}
\label{subsec:eval_backend_quality}

\begin{figure}
    \centering
    \includegraphics[width=\linewidth]{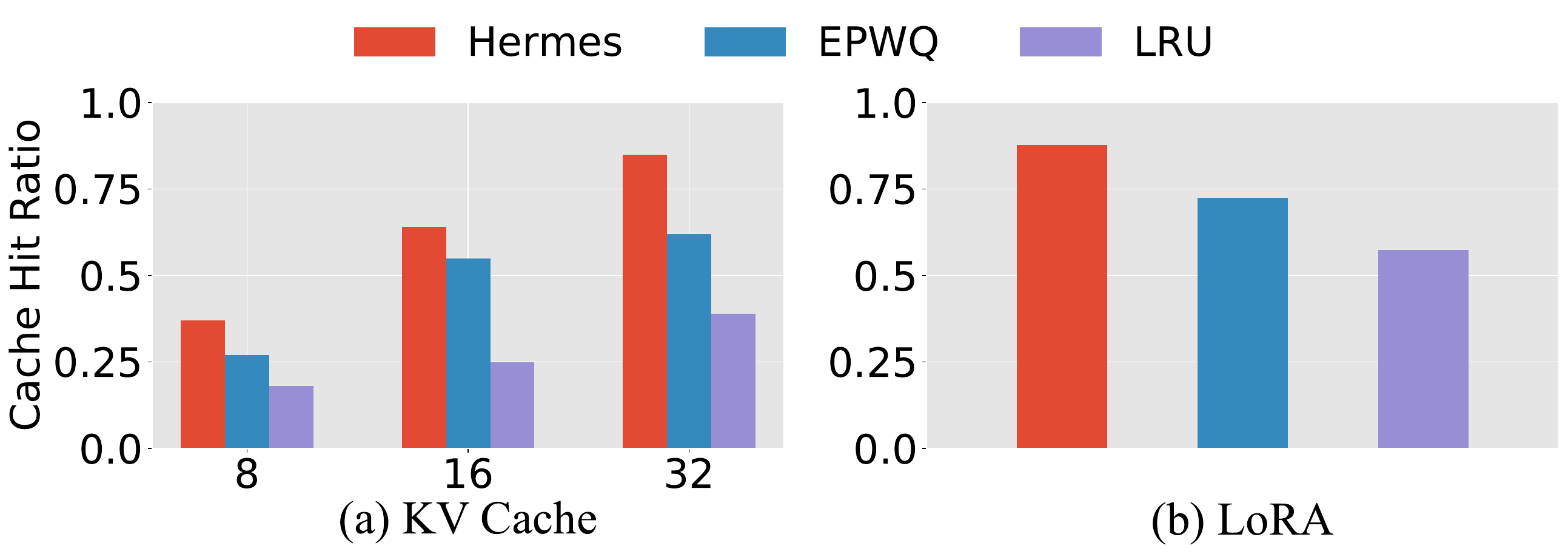}
            % \vspace{-0.12in}
    \caption{The cache hit ratio of (a) the KV Cache and (b) the LoRA adapter across different cache management strategies.}
            % \vspace{-0.1in}
    \label{fig:kvc_lora_chr}
\end{figure}

\phm{Effect of LLM backend prewarming.}
As explained in \Cref{subsec:provisioning}, prewarming the KV and LoRA cache can facilitate LLM serving.
% To evaluate the effectiveness of Hermes in KV cache management, 
To confirm that, we submit 500 applications to the A100 server within 15 minutes.
%We enable the KV cache to be evicted from GPU DRAM to CPU memory and also from the CPU memory to disk (whenever a layer is full), and loaded vice versa. 
We adopt two cache management baselines: LRU and Evict/Prefetch-Waiting-Queue (EPWQ)~\cite{gao2024cost,sheng2023s}. 
LRU conducts reactive swapping with prewarming, and EPWQ prewarms the KV cache only when the request is already in the waiting queue---which is often too late.
%means to determine the priority of a cached content based on its request order in the local waiting queue;
% conduct cache eviction based on the pending request queue (adopted in CachedAttention~\cite{gao2024cost}); 
%yet, during the inter-stage gap of an application, since the downstream request is not in the queue, its desired KV cache would be of the lowest priority.
Under each method, we measure the overall cache hit ratio (i.e., KV cache be well warmed when the request comes) against different cache sizes (8GB, 16GB, and 32GB). 
As shown in Fig.~\ref{fig:kvc_lora_chr}(a), Hermes consistently achieves the highest cache efficiency, improving the overall cache hit ratio by up to $1.11\times$ ($0.33\times$) compared to LRU (EPWQ).
It reduces the average ACT by 18\% (6\%) compared to LRU (EPWQ).
%Fig.~\ref{fig:kvc_exp}(b) further depicts the average ACT and makespan under each scheme in the case with 32GB CPU memory.
%It shows that Hermes can reduce the average ACT by 18\% (6\%) compared to LRU (EPWQ).

We then evaluate the benefit to prewarm the LoRA cache.
%\phm{Queue management of LoRA cache.} 
% In terms of KV cache and LoRA, we conducted a comprehensive load testing on an A100 to validate the performance of Clairvoyant cache management. We compared Hermes with three representative cache management strategies: vLLM, LRU, and Evict/Prefetch on Waiting Queue (EPWQ). As one of the most popular LLM inference frameworks, vLLM only stores free prefix KV cache in the remaining GPU space and does not utilize memory and disk as temporary storage. In contrast, we expand storage to include memory and disk, managing it with LRU, which we denote as one of our baselines. EPWQ is an enhanced cache management strategy based on information from the upper scheduler, actively evicting and prefetching the corresponding KV cache and LoRA adapters based on the requests waiting in the queue, reflecting the caching management concepts from the latest works, S-LoRA~\cite{sheng2023s} and CacheAttention~\cite{gao2024cost}. 
%To evaluate Hermes performance in LoRA caching, we compare Hermes with EPWQ, LRU, and No-Cache (i.e., not storing LoRA in host memory, which is also the default setting of vLLM). 
We set \texttt{max-loras} (the maximum number of parallel processes per iteration) to 10, \texttt{max-cpu-loras} (the number of LoRA caches that can be stored in CPU memory) to 20, and use totally 200 LoRA adapters for LLaMA-7B with a rank of 8. 
We submit a 25-minute workload containing 1,000 applications, with each application randomly assigned to one of the LoRA adapters (all requests of an application use the same LoRA).
Fig.~\ref{fig:kvc_lora_chr}(b) shows the cache hit ratio also respectively under LRU, EPWQ and Hermes. %. load latency and miss count under each scheme.
Compared to the second best (EPWQ), Hermes can increase the cache hit ratio by 21.1\%. %\yfliu{reduce the number of misses by 55.2\%}. 
%reduce the LoRA load latency by 45\%, and reduce the number of LoRA misses by 55.2\%.

\begin{figure*}
    \centering
    \includegraphics[width=\linewidth]{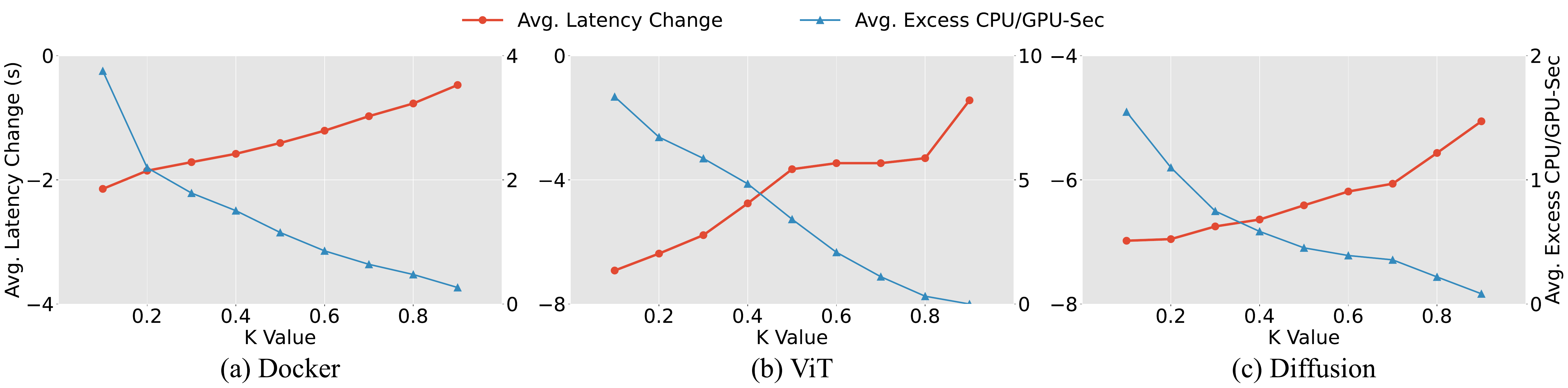}
    \caption{Benefit and cost with different prewarming aggressiveness levels ($K$) for non-LLM backends. 
    {A lower K value indicates more aggressive prewarming.}
    }
    \label{fig:prewarming-tradeoff}
\end{figure*}

\phm{Effect of non-LLM backend prewarming.}
To evaluate our prewarming strategy in non-LLM backends, we selected two applications, CG and PE, which respectively demands Docker and DNN backends. 
%Recall that in \Cref{subsec:provisioning} there is a trade-off in prewarming aggressiveness, and here we change the expected-prewarming-effectiveness $K$ to different values. 
We separately run each application with varying levels of prewarming aggressiveness (i.e., the expected prewarming effectiveness hyperparameter, $K$). Fig.~\ref{fig:prewarming-tradeoff} illustrates the average latency reduction and resource wastage due to prewarming, which verifies the effectiveness of prewarming in non-LLM backends (for DNN backends, we only analyze the PE applications calling a ViT or Diffusion model).
%(the extra time Docker containers or models remain active) compared to strategy without prewarming for each backend. 
Moreover, Fig.~\ref{fig:prewarming-tradeoff} also shows that a smaller $K$ can in general yield faster completion, yet at the cost of higher resource wastage---consistent with our analysis in \Cref{subsec:provisioning}.
%latency reduction but also raises the risk of resource wastage.

\subsection{Overhead Analysis}
\label{subsec:overhead}
% \begin{figure}[t]
%     \centering
%     \includegraphics[width=0.5 \linewidth]{figures/evaluation_overhead.pdf}
%     \caption{The median policy runtime for Hermes scheduler across different application arrive rates. The error bars indicate the 25$^{th}$ and 75$^{th}$ percentiles.}
%     \label{fig:overhead}
% \end{figure}

\begin{figure}
    \centering
    \subfloat[The policy runtime of Gittins over different application arrive rates. ]{
        \label{fig:policy_runtime}
        \centering
        \includegraphics[width=0.45\linewidth]{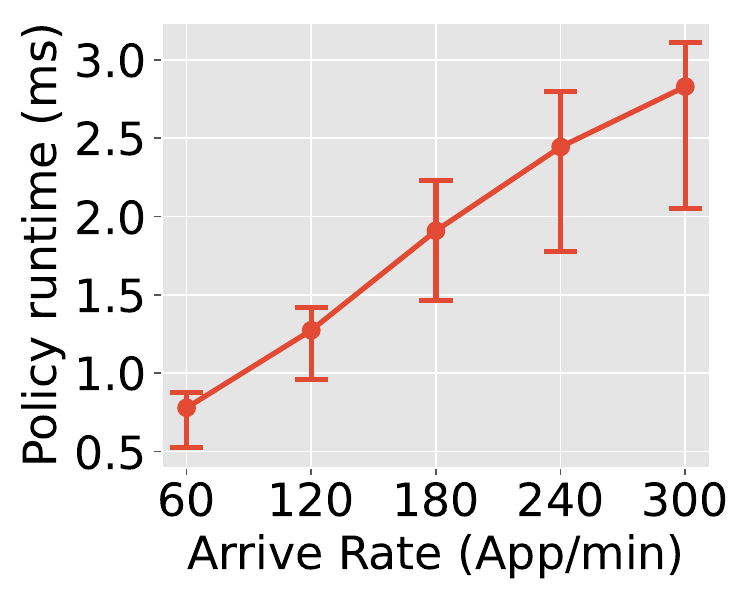}
    }
    \hfill
    \subfloat[How Gittins policy runtime and ACT vary with the bucket numbers. ]{
        \label{fig:policy_runtime_bucket}
        \centering
        \includegraphics[width=0.45\linewidth]{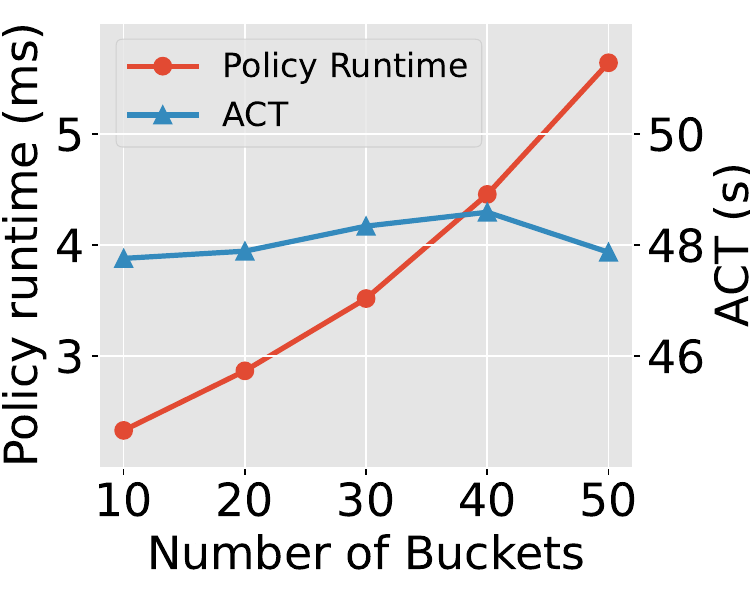}
    }
    \caption{Scheduling policy runtime for Hermes under various application arriveal rates using proportionally sized MoonCake traces. %The overhead of the Hermes scheduler during each scheduling round and the updates triggered upon application completion.
    }
    \label{fig:overhead}
\end{figure}

The main overhead of Hermes lies in determining the application scheduling priority with Gittins, which is iteratively executed once after a time length of the bucket size (buckets are used in describing the demand distribution as in Fig.~\ref{fig:token_usage}).
In Fig.~\ref{fig:policy_runtime}, we measure the average policy runtime of Gittins under different arrival rates (indicating different scheduling scales).
It shows that the scheduling overhead is indeed quite small in each case (less than 3 ms).
This is because our PDGraph models are indeed not large and can be efficiently processed.
Further in Fig.~\ref{fig:policy_runtime_bucket}, we measure how the Gittins policy runtime varies with the bucket number
% (duration range divided by bucket size)
.
It shows that using more buckets to describe the demand distribution almost linearly increases the policy runtime---but does not help improve the scheduling performance.

% Hermes introduces two primary sources of overhead: the first is the computation of Gittins rank during each scheduling round, and the second is the update of the history distribution information and the Gittins cache upon application completion.
% To evaluate its performance, we measure the scheduling overhead under various application arrival rates during each scheduling round, as well as the Gittins cache update time across different numbers of buckets.
% It is worth noting that in our evaluation, even under the most densely packed workload, the application arrival rate does not exceed 1 app/s. Additionally, our ablation experiments~\S\ref{subsec:ablation} confirm that setting the number of buckets for Gittins rank to 10 is sufficient to achieve an accurate priority estimation.
% As shown in Fig.~\ref{fig:overhead}, Hermes' computational overhead remains below 1 ms during each scheduling round and below 2 ms upon each application completion. 
% This low overhead solidifies Hermes' ability to deliver both fast execution speed and high throughput.

%!TEX root = main.tex
\section{Related Work and Discussions}
\label{sec:related}

\phm{Accelerating individual LLM requests.}
    Many works seek to accelerate individual LLM requests in a series of aspects.
    In \emph{operator implementation} aspect, the FlashAttention~\cite{dao2022flashattention} and FlashDecoding~\cite{hong2024flashdecoding} style algorithms are commonly adopted to realize efficient LLM inference by integrating IO awareness. 
    In \emph{model deployment} aspect, model-quantization~\cite{frantar2022gptq,kim2023squeezellm} and paged-attention~\cite{kwon2023efficient} methods have been proposed to maximize backend utilization.
    In \emph{inference algorithm} aspect, continuous batching~\cite{yu_orca_2022} and speculative decoding methods~\cite{miao2023specinfer,cai2024medusa} are also applied to improve the inference throughput by accelerating the decoding process.
    These works are orthogonal to us. 
    %\yfliu{'Adaptability to Adavanced Architecture' has been mentioned here.}

\phm{Scheduling multiple LLM requests.}
    Some recent works do notice the need to optimize scheduling performance at the request level.
    FastServe~\cite{wu2023fast} adopts multi-level feedback queue for LLM inference serving (yet at the cost of relatively high preemption overhead); Llumnix~\cite{sun2024llumnix} and LoongServe~\cite{wu2024loongserve} enable runtime request migration to improve load-balancing and isolation.
    In the meantime, some other works seek to
    %to predict the output length of individual requests~\cite{jin2023s,qiu2024efficient,shahout2024don} and 
    adopt SJF-like scheduling algorithms~\cite{shahout2024don,hu2024inference,10.1145/3698038.3698523}. 
    While those methods can reduce the average completion time of requests, they are application-agnostic and, as explained in \Cref{subsec:limitation}, fail to yield fast application completion. %\chen{infercept}
    Meanwhile, Hermes can be easily extended to facilitate multi-backend routing, enabling interference avoidance when co-locating requests on the same engine. For example, with the input/output length information recorded in PDGraph, it is possible to pack demand-complementary requests (e.g., those with long outputs which are memory-intensive and those with short outputs which are compute-intensive) onto a specific engine.

\phm{Job demand prediction.}
Demand prediction is important for efficient job scheduling. 
In traditional fields, a series of research works have been proposed to make accurate performance prediction with testbed profiling~\cite{alipourfard2017cherrypick,zheng2020nfv} or mathematical modeling~\cite{peng2018optimus,hu2023lucid}. 
Yet LLM workloads exhibit distinct demand uncertainty due to its generative manner, which is never captured before.
Some recent methods have been proposed to predict the output token length of an LLM request---with LLMs themselves~\cite{jin2023s,zheng2024response,shahout2024don,qiu2024efficient}, at the cost of lengthy model fine-tuning processes.
%, or with retrieval-based methods~\cite{zhao2024alise,10.1145/3698038.3698523}.
Moreover, none of those methods make demand prediction from the application point of view; blind to the task dependencies, their prediction accuracy is low, and in the meantime they cannot support speculative prewarming of diverse backends.

\phm{Extensibility of Hermes in advanced inference architectures.}
Prefill-decode-separation~\cite{zhong_distserve_2024,agrawal_taming_2024} has recently emerged as a popular LLM deployment paradigm that can avoid inter-request interferences.
Hermes can work smoothly with such disaggregated inference architectures. On the one hand, the queuing algorithm in Hermes works at the global level, and the resultant priority applies to all the backends; on the other hand, model or KV cache prewarming remains a common need for each prefill or decode instance under the PD-disaggregation architecture.
Meanwhile, Hermes can also be easily extended to facilitate multi-backend routing~\cite{hu2024inference,nie2024aladdin}, enabling interference avoidance when co-locating requests on the same engine. For example, with the input/output length information recorded in PDGraph, it is possible to pack demand-complementary requests (e.g., those with long outputs which are memory-intensive and those with short outputs which are compute-intensive) onto a specific engine. We will explore such functionalities with Hermes in the future.

\section{Conclusion}
\label{sec:conlucsion}

% In this paper, we propose Hermes, an efficient serving system for LLM applications based on the coinference abstraction.
% Hermes models a coinference as a list of stages, and depicts the dynamic resource demand in each aspect as a distribution.  
% It further adopts the Gittins policy to optimize inter-coinference scheduling performance supporting diverse objectives.
% For fast execution of each LLM application, Hermes adopts clairvoyant resource provisioning based on the profiled coinference demand on each backend.
% Experimental results on popular workloads show that Hermes can remarkably improve the serving efficiency of LLM applications.

In this paper, we propose Hermes, an efficient serving system designed for LLM applications. 
Hermes employs a probabilistic demand graph (PDGraph) to model the resource demands of LLM applications. 
Using the PDGraph, Hermes adopts the Gittins policy in queuing management to minimize the average application completion time, and adopts the LSTF algorithm to maximize the deadline satisfactory ratio for the cases with explicit deadlines. %which are propagated to each backend and also extended to for cases with deadlines.
%optimize the scheduling performance of LLM applications with probabilistic demands and propagates priorities across multiple backends to enable  coordinated scheduling. 
Hermes also leverages the PDGraph model to determine when and what backend to prewarm. %the desired type and time for backend prewarming. % prewarm the cold backends. 
Experimental results on popular workloads show that Hermes can remarkably improve the serving efficiency of LLM applications, attaining an improvement of over 70\%.

\bibliographystyle{plain}
\bibliography{main}
%% Bibliography
%\bibliography{bibfile}

%\input{appendix}

\end{document}